\documentclass[final,3p,times]{elsarticle}

\usepackage{amssymb, amsmath,framed,amsthm}

\journal{Journal of Computational Physics}

\usepackage{color,soul}
\usepackage{graphicx}

\usepackage[normalem]{ulem}

\usepackage[colorlinks=true,linkcolor=blue]{hyperref}
\newcommand{\p}{\partial}
\newcommand{\f}[2]{\frac{#1}{#2}}
\newcommand{\mr}[1]{\mathrm{#1}}
\newcommand{\lf}{\left}
\newcommand{\ri}{\right}

\newcommand{\dd}[2]{\frac{\rmd#1}{\rmd#2}}
\newcommand{\pp}[2]{\frac{\p #1}{\p #2}}

\newcommand{\de}[2]{\frac{\delta #1}{\delta #2}}

\newcommand{\nbl}{\nabla}
\newcommand{\para}{{||}}
\newcommand{\unit}[1]{\hat{\bm{#1}}}

\makeatletter
\newcommand{\vast}{\bBigg@{4}}
\newcommand{\Vast}{\bBigg@{5}}

\newcommand\agk{{\tt AstroGK}}
\newcommand\gs{{\tt GS2}}
\newcommand{\bm}[1]{{\boldsymbol {\mathrm #1}}} 
 
\newcommand{\zhat}{\unit{z}}

\makeatother

\newcommand{\calM}{\mathcal{M}}
\newcommand{\calN}{\mathcal{N}}

\DeclareMathAlphabet{\mathpzc}{OT1}{pzc}{m}{it}

\newcommand{\pzcG}{\mathpzc{G}}

\newcommand{\rma}{\mathrm{a}}

\newcommand{\rmd}{\mathrm{d}}
\newcommand{\rme}{\mathrm{e}}

\newcommand{\rmi}{\mathrm{i}}

\newcommand{\rmA}{\mathrm{A}}

\newcommand{\sfA}{\mathsf{A}}
\newcommand{\sfB}{\mathsf{B}}
\newcommand{\sfC}{\mathsf{C}}
\newcommand{\sfD}{\mathsf{D}}
\newcommand{\sfE}{\mathsf{E}}
\newcommand{\sfF}{\mathsf{F}}
\newcommand{\sfG}{\mathsf{G}}
\newcommand{\sfH}{\mathsf{H}}
\newcommand{\sfI}{\mathsf{I}}
\newcommand{\sfJ}{\mathsf{J}}
\newcommand{\sfK}{\mathsf{K}}
\newcommand{\sfL}{\mathsf{L}}
\newcommand{\sfM}{\mathsf{M}}

\newcommand{\sfP}{\mathsf{P}}
\newcommand{\sfQ}{\mathsf{Q}}

\newcommand{\sfT}{\mathsf{T}}

\newcommand{\sfZ}{\mathsf{Z}}

\begin{document}

\begin{frontmatter}


\title{A Hybrid Gyrokinetic Ion and Isothermal Electron Fluid Code for Astrophysical Plasma}
\author[Oxford]{Y.~Kawazura\corref{cor}}
\ead{yohei.kawazura@physics.ox.ac.uk}
\cortext[cor]{Corresponding author}

\author[Oxford,CCFE]{M.~Barnes}

\address[Oxford]{Rudolf Peierls Centre for Theoretical Physics, University of Oxford, Oxford OX1 3NP, United Kingdom}
\address[CCFE]{Culham Centre for Fusion Energy, Culham Science Centre, Abingdon OX14 3DB, United Kingdom}


\author{}

\address{}

\begin{abstract}
  This paper describes a new code for simulating astrophysical plasmas that solves a hybrid model composed of gyrokinetic ions (GKI) and an isothermal electron fluid (ITEF) [A. Schekochihin et al., Astrophys. J. Suppl. \textbf{182}, 310 (2009)].
  This model captures ion kinetic effects that are important near the ion gyro-radius scale while electron kinetic effects are ordered out by an electron--ion mass ratio expansion.
  The code is developed by incorporating the ITEF approximation into \agk, an Eulerian $\delta f$ gyrokinetics code specialized to a slab geometry [R. Numata et al., J. Comput. Phys. \textbf{229}, 9347 (2010)]. 
  The new code treats the linear terms in the ITEF equations implicitly while the nonlinear terms are treated explicitly.
  We show linear and nonlinear benchmark tests to prove the validity and applicability of the simulation code.
  Since the fast electron timescale is eliminated by the mass ratio expansion, the Courant--Friedrichs--Lewy condition is much less restrictive than in full gyrokinetic codes; the present hybrid code runs $\sim 2\sqrt{m_\rmi/m_\rme} \sim 100$ times faster than \agk\ with a single ion species and kinetic electrons where $m_\rmi/m_\rme$ is the ion--electron mass ratio.
  The improvement of the computational time makes it feasible to execute ion scale gyrokinetic simulations with a high velocity space resolution and to run multiple simulations to determine
  the dependence of turbulent dynamics on parameters such as electron--ion temperature ratio
  and plasma beta.
\end{abstract}


\end{frontmatter}



\section{Introduction}\label{s:intro}
Understanding the thermodynamic properties of hot and dilute plasma is essential to advancing the study of astrophysics. 
This endeavor is particularly challenging because many astrophysical systems are in a weakly collisional state, where the collisional mean free path is comparable to or larger than the system size. 
Consequently, widely-used fluid models such as magnetohydrodynamics (MHD) are not suitable
for describing the microscale physics that determine thermodynamic equilibrium.  
Instead, we need to treat the plasma using kinetic theory, in which the distribution of particle positions and velocities evolves in a six--dimensional phase space. 
The computational cost required to conduct such six-–dimensional (6D) computations is enormous:
Even with the help of exascale computing, well-resolved 6D simulations are unlikely to be feasible in the near future.

Computational cost can be reduced significantly in magnetized plasma 
by adopting the gyrokinetic~\cite{Rutherford1968,Taylor1968,Catto1978,Frieman1982,Brizard2007} model.
A key ingredient of gyrokinetics is an assumed time scale separation created by the presence of a magnetic field;
the ion cyclotron timescale is taken to be much faster than the timescale for the fluctuations of interest, i.e. $\Omega_\rmi \gg \omega$ with the ion cyclotron frequency $\Omega_\rmi$ and the fluctuation frequency $\omega$.
This separation allows us to reduce the phase space from 6D to 5D --- 3D in position space and 2D in velocity space --- by averaging out the fast cyclotron motion. 
In the past few decades, gyrokinetics has been extensively used for studying microinstabilities and transport in magnetic confinement fusion devices~\cite{Dimits2000,Garbet2010}.
Recently, gyrokinetics has been highlighted as a powerful model in astrophysics as well~\cite{Howes2006ApJ,Schekochihin2009ApJ}.
While many of the past astrophysical studies via gyrokinetics have targeted the solar wind~\cite{Howes2008,Howes2011,TenBarge2012,Told2015,Navarro2016,Howes2017,Bourouaine2017,Klein2017}, gyrokinetics is also expected to be applicable to many other astrophysical objects, such as accretion flows, galaxy clusters, and interstellar media~\cite{Schekochihin2009ApJ}.


Although gyrokinetics improves the computational cost dramatically, it is still challenging to calculate 3D electromagnetic problems with high velocity space resolution.
A further simplification to the gyrokinetic model can be made by carrying out an asymptotic expansion
of the gyrokinetic-Maxwell system of equations in the smallness of the electron-ion mass ratio, $\sqrt{m_\rme/m_\rmi} \sim 0.02$, with $m_\rme$ and $m_\rmi$ the electron and proton masses, respectively.  This limits the applicability of the model to spatial scales in the plane perpendicular to the mean magnetic
field that are comparable to or larger than the ion Larmor radius.  However, it also eliminates the fast time scale
associated with the electron's thermal speed and reduces the electron gyrokinetic equation to a set of
fluid equations in which the electron temperature is constant along the mean magnetic field~\cite{Snyder1999,Snyder2001a,Schekochihin2009ApJ}.  If one further assumes that the electron temperature does not vary across the mean magnetic field -- as might be the case when the field is tangled -- then one obtains the isothermal electron fluid (ITEF) model~\cite{Schekochihin2009ApJ}.

Coupling the ITEF with gyrokinetic ions (GKI) leads to a hybrid model that reduces
computational cost by a factor of $\sim \sqrt{m_\rme/m_\rmi}$ relative to full gyrokinetic model (FGK). 
\footnote{
We note that the mass ratio expansion approach may be also applied for the full Vlasov--Maxwell system to derive fully kinetic ions and an electron fluid hybrid model, which is capable of capturing the high frequency (faster than ion cyclotron frequency) dynamics for ions.
This model has a long history both in fusion~\cite{Sgro1976,Byers1978,Hewett1978,Hewett1980,Harned1982,Matthews1994} and astrophysical contexts~\cite{Liewer2001,Hellinger2003,Hellinger2005, Kunz2014,Kunz2016,Cerri2017,Groselj2017}.
In terms of the computational algorithm, the particle--in--cell method is employed to solve ion motion in most of the cases (except for Ref.~\cite{Valentini2007} which solves the Vlasov equation in an Eulerian description).
One may see the GKI/ITEF hybrid model not only as a massless electron reduction of FGK but also as a low frequency reduction of the Vlasov ion and electron fluid hybrid model.
}
While such a hybrid model with GKI and an ITEF has a long history in magnetic confinement fusion~\cite{Lin2001fluid,Chen2001,Snyder2001b,Parker2000,Abel2013,Hinton2003,Hager2017}, it has not been applied to astrophysical plasmas.
Moreover, unlike magnetic confinement fusion plasmas, many astrophysical plasmas have a plasma pressure comparable to and possibly much larger than the magnetic pressure; i.e.,
$\beta_\rmi = 8\pi p_\rmi/B^2 \sim 1-100$, with $p_\rmi$ the ion pressure and $B$ the magnetic field amplitude.
This makes simulations rather cumbersome and increases the need for computational savings such as those provided by a hybrid code.

Therefore, it is valuable to develop a fluid--gyrokinetic hybrid code specialized for astrophysical studies. 
In this work, we develop a simulation code that solves the gyrokinetic ion and isothermal electron fluid (GKI/ITEF) equations~\cite{Schekochihin2009ApJ} by implementing a new algorithm in \agk~\cite{Numata2010JCP}, a local, Eulerian, $\delta f$ gyrokinetics code developed for astrophysical studies.  \agk~\cite{Numata2010JCP} is based on the magnetic confinement fusion code \gs~\cite{Kotschenreuther1995CPC,Dorland2000} but is optimized to treat plasmas with a straight, homogeneous mean magnetic field.  Electrostatic simulations with \agk\ have been used to study the entropy cascade in phase space~\cite{Tatsuno2009,Tatsuno2010,Tatsuno2012,Plunk2011}, and
electromagnetic simulations have been used to study magnetic reconnection~\cite{Numata2011,Kobayashi2014,Numata2015,Zocco2015} in addition to the research on solar wind turbulence (reference listed above).
To implement our hybrid model, we have modified the \agk\ algorithm to solve the ITEF equations coupled to the GKI-Maxwell's system of equations.
Since the most computationally intensive part of the hybrid code -- the solution of the ion gyrokinetic equation -- is unchanged, it retains the excellent parallel performance of \agk.

The outline of this paper is as follows.
Section~\ref{s:model} presents a set of equations for the gyrokinetic model followed by the GKI/ITEF hybrid model.  In Section~\ref{s:algorithm}, we describe the numerical algorithm for the hybrid code.
We first briefly show the time integration algorithm adopted in \agk.
We then show the detailed algorithm for solving the ITEF.
In Section~\ref{s:CFL improvement}, we estimate the computational savings of the hybrid code relative to FGK codes.  
Section~\ref{s:tests} presents the results of linear and nonlinear benchmark tests for code verification.
It is found that the nonlinear test not only demonstrates the validity of the code but also reveals a non--trivial result which could be the starting point of a future study.
In Section~\ref{s:conclusion}, we close with a summary of the paper.

\section{Model equations}\label{s:model}
We start by presenting the FGK system of equations. 
Let us consider a homogeneous plasma immersed in magnetic field $\bm{B}_0 = B_0\zhat$.
In astrophysical systems, $\bm{B}_0$ may be assumed to be constant straight field at microscale.
In $\delta f$ gyrokinetics, a particle distribution function $f_s$ for species $s$ is split into the mean and fluctuating parts: 
\begin{equation}
  f_s(\bm{r},\bm{v}, t) = F_{s} + \delta f_s = \lf( 1 - \f{q_s \phi(\bm{r})}{T_{s}} \ri)F_{s}(v) + h_s(t,\bm{R}_s,v_\para,v_\perp)
  \label{e:def h}
\end{equation}
where $F_{s} = n_{s}(m_s/2\pi T_{s})^{3/2}\exp(-m_s v^2/2T_{s})$ is the mean distribution function assumed to be a Maxwellian with equilibrium density $n_{s}$ and temperature $T_{s}$, $\delta f_s$ is the fluctuating distribution function, $\phi$ is the fluctuating electrostatic potential, $h_s(t,\bm{R}_s,v_\para,v_\perp)$ is the non-Boltzmann part of $\delta f_s$, $\bm{R}_s = \bm{r} + \bm{v}\times\zhat/\Omega_s$ is guiding center position, and $\para$ ($\perp$) denotes the direction parallel (perpendicular) to $\zhat$.
In the present study, we assume $n_{s}$ and $T_{s}$ are homogeneous in space.
Electromagnetic (EM) fields are expressed in terms of the scalar and vector potentials, $\phi$ and $\bm{A}$, as 
\begin{equation}
  \delta\bm{B} = \nbl_\perp A_\para\times\zhat + \delta B_\para\zhat, \quad \bm{E} = -\nbl\phi - \f{1}{c}\pp{\bm{A}}{t}. 
  \label{e:def B & E}
\end{equation}
Substituting the forms for the distribution function and EM fields given by \eqref{e:def h} and \eqref{e:def B & E} into the Maxwell--Boltzmann system of equations and applying the gyrokinetic ordering, $\epsilon \sim k_\para/k_\perp \sim \omega/\Omega_s \sim \delta f_s/F_{s} \sim \delta B/B_0$, we obtain the gyrokinetic equation~\cite{Howes2006ApJ} 
\begin{equation}
  \pp{h_s}{t} + v_\para\pp{h_s}{z} + \f{c}{B_0} \lf\{ \langle \chi \rangle_{\bm{R}_s} ,h_s \ri\} = \f{q_s}{T_s}\pp{\lf< \chi \ri>_{\bm{R}_s}}{t}F_s + \lf< C[h_s] \ri>_{\bm{R}_s},
  \label{e:dhdt in real space}
\end{equation}
and Maxwell's equations, viz., the quasi-neutrality condition, the parallel and perpendicular Amp{e}re's law,
\begin{subequations}
\begin{align}
  &\sum_s \f{q_s^2 n_{s}}{T_{s}} \phi = \sum_s q_s \int\rmd^3 \bm{v}\, \lf< h_s \ri>_\bm{r},
  \label{e:quasineutrality in real space} \\
  &-\f{c}{4\pi}\nbl_\perp^2A_{\para} = \sum_s q_s \int\rmd^3 \bm{v}\, v_\para \lf< h_s \ri>_\bm{r}, 
  \label{e:para Ampere in real space} \\
  &\f{c}{4\pi}\nbl_\perp\delta B_{||} = \sum_s T_{s}\int\rmd^3 \bm{v}\, \lf< (\zhat\times\bm{v}_\perp)h_s \ri>_\bm{r},
  \label{e:perp Ampere in real space}
\end{align}
\end{subequations}
where $\chi = \phi - \bm{v}\cdot\bm{A}/c$ is the gyrokinetic potential, $\lf< \cdots \ri>_{\bm{R}_s}$ and $\lf< \cdots \ri>_{\bm{r}}$ are the gyro--averages at fixed $\bm{R}_s$ and $\bm{r}$, respectively, $C[h_s]$ is a linearized collision operator that includes pitch angle scattering and energy diffusion satisfying conservation properties~\cite{Abel2008,Barnes2009}, and
$\lf\{ \cdots\, ,\, \cdots \ri\}$ is the Poisson bracket defined by
\begin{equation}
  \lf\{ a , b \ri\} = \zhat\cdot\pp{a}{\bm{R}_s}\times\pp{b}{\bm{R}_s}.
  \label{e:Poisson bracket}
\end{equation}

The assumed homogeneity of $F_s$ admits periodic solutions to these equations.
We thus Fourier transform the fields in the plane perpendicular to $\zhat$:
\begin{subequations}
\begin{align}
  h_s(\bm{R}_s, v_\para, v_\perp, t) &= \sum_{\bm{k}_\perp} h_{s\bm{k}_\perp}(z,v_\para, v_\perp, t)\, \rme^{\rmi\bm{k}_\perp\cdot\bm{R}_s} \\
  \phi(\bm{r}, t) &= \sum_{\bm{k}_\perp} \phi_{\bm{k}_\perp}(z,t)\, \rme^{\rmi\bm{k}_\perp\cdot\bm{r}}. 
\end{align}
\end{subequations}
The gyrokinetic equation in terms of Fourier component is
\begin{multline}
  \f{\p g_{s\bm{k}_\perp}}{\p t} + v_\para\pp{g_{s\bm{k}_\perp}}{z} + v_\para\pp{}{z}\lf( J_0(a_s)\f{q_s\phi_{\bm{k}_\perp}}{T_{s}} + \f{2v_\perp^2}{v_{\mr{th}s}^2}\f{J_1(a_s)}{a_s}\f{\delta B_{||\bm{k}_\perp}}{B_0} \ri)F_{s} + \f{c}{B_0} \lf\{ \langle \chi \rangle_{\bm{R}_s} ,h_s \ri\}_{\bm{k}_\perp} \\
  = -\f{q_sF_{s}}{T_{s}}\f{v_\para}{c}J_0(a_s) \f{\p A_\para}{\p t} + C_\mr{GK}\lf[h_{s\bm{k}_\perp}\ri],
  \label{e:dgdt}
\end{multline}
where $g_{s\bm{k}_\perp}$ is a complementary distribution function defined by
\begin{equation}
  g_{s\bm{k}} = h_{s\bm{k}} - \f{q_s F_{s}}{T_{s}}\lf( J_0(a_s)\phi_{\bm{k}_\perp} + \f{J_1(a_s)}{a_s}\f{2 v_\perp^2}{v_{\mr{th}s}^2}\f{T_{s}}{q_s}\f{\delta B_{\para\bm{k}_\perp}}{B_0} \ri),
  \label{e:def g}
\end{equation}
$J_n(a_s)$ is the Bessel function of the first kind with the argument $a_s = k_\perp v_\perp/\Omega_s$,
$\lf\{ \cdots\, ,\, \cdots \ri\}_{\bm{k}_\perp}$ is the Fourier transform of the Poisson bracket,
$v_{\mr{th}s} = \sqrt{2T_{s}/m_s}$ is the thermal speed,
and $C_\mr{GK}$ is the Fourier component of the gyro--averaged collision operator, i.e., $\lf< C[h] \ri>_{\bm{R}_s} = \sum_{\bm{k}_\perp}\rme^{\rmi\bm{R}_s\cdot\bm{k}_\perp}C_\mr{GK}[h_{\bm{k}_\perp}]$.
Here the arguments of the Poisson bracket are evaluated in real space because we treat the nonlinear term by the pseudo-spectral method in the simulation code.
The Fourier components of Maxwell's equations are
\begin{subequations}
\begin{align}
  &\sum_s \f{q_s^2 n_{s}}{T_{s}} [1 - \Gamma_{0}(\alpha_s)] \phi_{\bm{k}_\perp} - \sum_s q_s n_{s} \Gamma_{1}(\alpha_s) \f{\delta B_{||\bm{k}_\perp}}{B_0} = \sum_s q_s \int\rmd^3 \bm{v}\, J_0(a_s)g_{s\bm{k}_\perp},
  \label{e:quasineutrality FGK} \\
  &\f{ck_\perp^2}{4\pi}A_{\para\bm{k}_\perp} = \sum_s q_s \int\rmd^3 \bm{v}\, v_\para J_0(a_s)g_{s\bm{k}_\perp}, 
  \label{e:para Ampere FGK} \\
  &\f{B_0}{4\pi}\delta B_{||\bm{k}_\perp} + \sum_s q_s n_{s} \Gamma_{1}(\alpha_s)\phi_{\bm{k}_\perp} + \sum_s n_{s} T_{s} \Gamma_{2}(\alpha_s)\f{\delta B_{||\bm{k}_\perp}}{B_0} = -\sum_s T_{s}\int\rmd^3 \bm{v}\, \f{2v_\perp^2}{v_{\mr{th}s}^2}\f{J_1(a_s)}{a_s} g_{s\bm{k}_\perp},
  \label{e:perp Ampere FGK}
\end{align}
\end{subequations}
where $\Gamma_n(\alpha_s)$ is defined by 
\begin{equation}
  \Gamma_0(\alpha_s) = I_0(\alpha_s)\rme^{-\alpha_s}, \quad \Gamma_1(\alpha_s) = [I_0(\alpha_s) - I_1(\alpha_s)]\rme^{-\alpha_s}, \quad \Gamma_2(\alpha_s) = 2\Gamma_1(\alpha_s),
\end{equation}
with the $I_n$ modified Bessel functions of the first kind and $\alpha_s = k_\perp^2\rho_s^2/2$.
\agk\ solves \eqref{e:dgdt} and \eqref{e:quasineutrality FGK}--\eqref{e:perp Ampere FGK}.

The GKI/ITEF equations are obtained by imposing two approximations to FGK, namely, massless electron ($\sqrt{m_\rme/m_\rmi} \ll 1$) and isothermal electron closure ($\delta T_\rme = 0$)~\cite{Schekochihin2009ApJ}.
Expansion of \eqref{e:dgdt} for electrons and \eqref{e:quasineutrality FGK}--\eqref{e:perp Ampere FGK} with respect to the small parameter $\sqrt{m_\rme/m_\rmi}$ give a set of equations.
One of the resulting equations, $\unit{b}\cdot\nbl\delta T_\rme = 0$ with the total magnetic field direction $\unit{b} = \zhat + \delta\bm{B}_\perp/B_0$, restricts the electron temperature fluctuation along magnetic field line to be constant.
We further assume that $\delta T_\rme = $ constant, as would be the case if the magnetic field lines are tangled.
By assuming $\delta T_\rme = 0$ (isothermal electron closure), the non--Boltzmann part of the perturbed electron distribution function is written by
\begin{equation}
  h_\rme^{(0)} = \lf( \f{\delta n_\rme}{n_\rme} - \f{e}{T_\rme}\phi \ri)F_\rme,
  \label{e:h_e^0}
\end{equation}
where the superscript $(0)$ indicates zeroth order in $\sqrt{m_\rme/m_\rmi}$.
By substituting $h_\rme^{(0)}$ to the electron gyrokinetic equation, one obtains the fluid dynamical equations for perturbed electron density $\delta n_\rme$ and parallel electron flow speed $u_{\para\rme}$.
In the resulting set of equations, the electron kinetic effects, which are primary effective at the electron gyro--radius scale, are neglected~\cite{Schekochihin2009ApJ}. 
We consider a single ion species with charge $Ze$.
The electron gyrokinetic equation is replaced by two dynamical equations:
\begin{subequations}
\begin{align}
  & \pp{}{t}\lf( \f{\delta n_{\rme\bm{k}_\perp}}{n_\rme} - \f{\delta B_{||\bm{k}_\perp}}{B_0} \ri) + \f{c}{B_0}\lf\{\phi - \f{T_\rme}{e}\f{\delta n_\rme}{n_\rme},\, \f{\delta n_\rme}{n_\rme} - \f{\delta B_{||}}{B_0}\ri\}_{\bm{k}_\perp} + \pp{u_{||\rme\bm{k}_\perp}}{z} = \f{1}{B_0}\{A_{||},\, u_{||\rme}\}_{\bm{k}_\perp},
  \label{e:delta ne evo} \\
  & \pp{A_{||\bm{k}_\perp}}{t} + \f{c}{B_0}\lf\{\phi - \f{T_\rme}{e}\f{\delta n_\rme}{n_\rme},\, A_{||}\ri\}_{\bm{k}_\perp} + c\pp{\phi_{\bm{k}_\perp}}{z} - \f{cT_\rme}{e}\pp{}{z}\lf( \f{\delta n_{\rme\bm{k}_\perp}}{n_\rme} \ri) = 0,
  \label{e:A|| evo}
\end{align}
\end{subequations}
where the spatial gradient for the Poisson bracket is defined by the particle position $\bm{r}$.
The Maxwell's equations \eqref{e:quasineutrality FGK}--\eqref{e:perp Ampere FGK} are rewritten as
\begin{subequations}
\begin{align}
  & \f{\delta n_{\rme\bm{k}_\perp}}{n_\rme} = \lf[ \Gamma_0(\alpha_\rmi) - 1 \ri]\f{Ze\phi_{\bm{k}_\perp}}{T_\rmi} + \Gamma_1(\alpha_\rmi)\f{\delta B_{\para\bm{k}_\perp}}{B_0} + \f{1}{n_\rmi}\int\rmd^3\bm{v}\, J_0(a_\rmi)g_{\rmi\bm{k}_\perp},
  \label{e:quasineutrality} \\
  & u_{\para\rme\bm{k}_\perp} = -\f{ck_\perp^2}{4\pi en_\rme}A_{\para\bm{k}_\perp} + \f{1}{n_\rmi}\int\rmd^3\bm{v}\, v_\para J_0(a_\rmi) g_{\rmi\bm{k}_\perp},
  \label{e:para Ampere} \\
  & \f{Z}{\tau}\f{\delta n_{\rme\bm{k}_\perp}}{n_\rme} + \lf[ \f{2}{\beta_\rmi} + \Gamma_2(\alpha_\rmi) \ri]\f{\delta B_{||\bm{k}_\perp}}{B_0} = \lf[ 1 - \Gamma_1(\alpha_\rmi) \ri]\f{Ze\phi_{\bm{k}_\perp}}{T_\rmi} - \f{1}{n_\rmi}\int\rmd^3\bm{v}\,\f{2v_\perp^2}{v_\mr{thi}^2}\f{J_1(a_\rmi)}{a_\rmi} g_{\rmi\bm{k}_\perp},
  \label{e:perp Ampere}
\end{align}
\end{subequations}
where $\tau = T_\rmi/T_\rme$ and $\beta_\rmi = 8\pi n_\rmi T_\rmi/B_0^2$.

Let us remark that the structure of the GKI/ITEF equations is altered from that of the FGK equations in the following sense.
In FGK, the dynamical variables are ion and electron distribution functions, and EM fields are subsequently determined by Maxwell's equation from the advanced distribution functions;
in GKI/ITEF, $A_\para$ is a dynamical variable, and $u_{\para\rme}$ is determined by the advanced $A_\para$.

\subsection{Generalized energy balance law}\label{ss:energy balance}
The generalized energy for the FGK system is defined as
\begin{equation}
  W = \sum_s E_{f_s} + E_{B} = \sum_s\int\rmd^3\bm{r}\int\rmd^3\bm{v}\,\f{T_{s}\delta f_s^2}{2F_{s}} + \int\rmd^3\bm{r}\,\f{|\delta\bm{B}|^2}{8\pi}.
  \label{e:FGK energy}
\end{equation}
The energy balance law is given by~\cite{Howes2006ApJ,Schekochihin2009ApJ}
\begin{equation}
  \dd{W}{t} = -\int\rmd^3\bm{r}\,\bm{J}_a\cdot\bm{E} + \sum_s\int\rmd^3\bm{R}_s\int\rmd^3\bm{v}\,\f{T_{s}}{F_{s}}\lf< h_s C[h_s] \ri>_{\bm{R}_s},
  \label{e:FGK dW/dt}
\end{equation}
where $\bm{J}_a$ denotes external current drive (see Section~\ref{s:antenna}) and the second term describes collisional entropy generation;
thus $W$ is a constant of the motion in the absence of external power injection and collisions. 

The generalized energy for the GKI/ITEF system is defined by~\cite{Schekochihin2009ApJ}
\begin{equation}
  W = E_{f_\rmi} + E_{n_\rme} + E_{B} = \int\rmd^3\bm{r}\int\rmd^3\bm{v}\,\f{T_\rmi\delta f_\rmi^2}{2F_\rmi} + \int\rmd^3\bm{r}\,\f{n_\rme T_\rme}{2}\f{\delta n_\rme^2}{n_\rme} + \int\rmd^3\bm{r}\,\f{|\delta\bm{B}|^2}{8\pi},
  \label{e:GKI/ITEF energy}
\end{equation}
and its time rate of change obeys \eqref{e:FGK dW/dt} with the electron collision ignored.
When the system is 2D, i.e., $\p_z = 0$, there is an extra invariant for GKI/ITEF~\cite{Schekochihin2009ApJ}:
\begin{equation}
  I_\rme = \int\rmd^3\bm{r}\,\f{A_\para^2}{2}.
  \label{e:A_\para^2 stuff}
\end{equation}
The conservation of these invariants is a good code verification, especially for a nonlinear run (Section~\ref{ss:2D OT}).

\subsection{Hyperviscosity for ITEF}\label{ss:hyperviscosity}
As shown in Ref.~\cite{Schekochihin2009ApJ}, the energy that is cascaded from large scales is separated around the ion Larmor scale into ion entropy fluctuations and kinetic Alfv\'{e}n waves (KAW).
The energy in these two channels is independently cascaded to smaller scales, and then dissipated into thermal energy of ions and electrons respectively via collisions.
In the GKI/ITEF hybrid model, the ion dissipation route exists, but the electron dissipation route is eliminated.
Hence, an artificial dissipation mechanism is required to terminate the KAW cascade at the smallest scales.
Such an artificial dissipation must (i) give a negative definite term in the right hand side of the energy balance equation \eqref{e:FGK dW/dt} and (ii) be effective only at the smallest scales of the computational domain.
We modify \eqref{e:delta ne evo} to include dissipation as follows:
\begin{equation}
  \pp{}{t}\lf( \f{\delta n_{\rme}}{n_\rme} - \f{\delta B_{||}}{B_0} \ri) + \f{c}{B_0}\lf\{\phi - \f{T_\rme}{e}\f{\delta n_\rme}{n_\rme},\, \f{\delta n_\rme}{n_\rme} - \f{\delta B_{||}}{B_0}\ri\} + \pp{u_{||\rme}}{z} = \f{1}{B_0}\{A_{||},\, u_{||\rme}\} + \nu_h\, \rho_\rmi^{2n}\nbl_\perp^{2n}\lf( \f{\delta n_\rme}{n_{e}} - \f{e}{T_{\rme}}\phi \ri).
  \label{e:delta ne evo w/ hyper}
\end{equation}
The last term represents a hyperviscosity term where $\nu_h$ is the hyperviscosity coefficient and $n$ is a positive integer.
This hyperviscosity term corresponds to the velocity integral of the collision operator acting on $h_\rme^{(0)}$, which is estimated by (see Appendix~B.1 in Ref.~\cite{Schekochihin2009ApJ})
\begin{equation}
  \f{1}{n_\rme}\int \rmd^3\bm{v}\lf< h_\rme^{(0)} C[h_\rme^{(0)}] \ri>_{\bm{R}_\rme} \sim \sqrt{\f{m_\rme}{m_\rmi}}\nu_{\rmi\rmi}k_\perp^2 \rho_\rmi^2\lf( \f{\delta n_\rme}{n_\rme} - \f{e}{T_\rme}\phi \ri),
  \label{e:h_e^0 viscosity}
\end{equation}
with the ion--ion collision frequency $\nu_\mr{ii}$.
Whereas this should be ordered out as it is first order in $\sqrt{m_\rme/m_\rmi}$, we can make it effective only at the small scales by changing $\sqrt{m_\rme/m_\rmi}\nu_\mr{ii}k_\perp^2\rho_\rmi^2 \to \nu_h k_\perp^{2n}\rho_\rmi^{2n}$.
When the electrons have a Boltzmann response, i.e. $h_\rme^{(0)} = 0$, the hyperviscosity vanishes.
  This is consistent with the behavior of the exact Landau-Boltzmann collision operator, whose kernel includes the Maxwell-Boltzmann distribution.
Using \eqref{e:delta ne evo w/ hyper}, the energy balance equation \eqref{e:FGK dW/dt} is modified as
\begin{equation}
  \dd{W}{t} = -\int\rmd^3\bm{r}\,\bm{J}_a\cdot\bm{E} + \int\rmd^3\bm{r}\int\rmd^3\bm{v}\,\f{T_\rmi}{F_\rmi}\lf< h_\rmi C[h_\rmi] \ri>_{\bm{R}_\rmi} - \nu_h n_{\rme}T_{\rme}\int\rmd^3\bm{r}\lf| \rho_\rmi^n\nbl_\perp^n \lf( \f{\delta n_\rme}{n_{\rme}} - \f{e}{T_{\rme}}\phi \ri) \ri|^2.
  \label{e:FGK dW/dt with hyperviscosity}
\end{equation}
Manifestly, the final term is negative definite.

We note that $A_\para$ may not be damped by the hyperviscosity. 
However, we expect this would not cause a problem because the nonlinear term of $A_\para$ evolution equation in $k_\perp \rho_\rmi \gg 1$ limit, which is proportional to $\{A_\para, \phi\}$~\cite{Schekochihin2009ApJ}, is negligible when $\phi$ is sufficiently damped by the hyperviscosity.
In fact, this has been confirmed by the 3D driven turbulence simulation~\cite{Kawazura2017b}.
On the other hand, $A_\para$ can be directly damped by adding a term which is proportional to $-\nbl_\perp^{2n}u_{||\rme}$ to the right hand side of \eqref{e:A|| evo}.
This term plays a role of hyperresistivity.
The new hybrid code does not have the hyperresistivity option at the moment as we have not found it necessary to attain converged results. However, it could be trivially implemented in the code if needed.

\subsection{Normalization}\label{ss:normalization}
We impose the same normalizations as those employed in \agk~\cite{Numata2010JCP} 
\begin{align}
  &
  z = L_\para\hat{z}, \quad
  x = \rho_0\hat{x}, \quad
  t = \f{L_\para}{v_\mr{th0}}\hat{t}, \quad
  \bm{v} = v_\mr{th\rmi}\hat{\bm{v}}_\rmi = v_\mr{th0}\hat{v}_\mr{thi}\hat{\bm{v}}_\rmi, \quad
  m_\rmi = m_0\hat{m}_\rmi, \quad
  \nonumber\\
  &
  n_\rmi = n_{0}\hat{n}_\rmi, \quad
  T_\rmi = T_{0}\hat{T}_\rmi, \quad
  \beta_\rmi = \f{8\pi n_\rmi T_\rmi}{B_0^2} = \hat{\beta}_0\hat{n}_\rmi\hat{T}_\rmi, \quad
  \nonumber\\
  &
  \delta n_\rme = \f{\rho_0}{L_\para}n_{0}\widehat{\delta n}_\rme, \quad
  u_{\para} = \f{\rho_0}{L_\para}v_{\mr{th}0}\hat{u}_{\para}, \quad
  \phi = \f{\rho_0}{L_\para}\f{T_{0}}{e}\hat{\phi}, \quad
  \delta B_\para = \f{\rho_0}{L_\para}B_{0}\widehat{\delta B}_\para, \quad
  A_\para = \f{\rho_0}{L_\para}\f{c T_{0}}{v_\mr{th0}e}\hat{A}_\para, \quad
  \nonumber\\
  &
  g_\rmi = \f{\rho_0}{L_\para}F_\rmi\hat{g}_\rmi, \quad
  \f{F_\rmi}{n_\rmi} \rmd^3\bm{v} = \widehat{\rmd^3\bm{v}}_\rmi, \quad
  \nu_h = \lf( \f{1}{k_\mr{max}\rho_\rmi} \ri)^{2n}\f{v_\mr{th0}}{L_\para}\hat{\nu}_h
\end{align}
where the subscripts 0 denote the reference values, $v_\mr{th0} = \sqrt{2T_0/m_0}$ is the reference thermal speed, $L_\para$ is the parallel scale length, and $k_\mr{max}$ is the maximum perpendicular wave number.
The resulting normalized GKI/ITEF equations are
\begin{multline}
  \pp{\hat{g}_{\rmi\bm{k}_\perp}}{\hat{t}} + \sqrt{\f{\hat{T}_{\mr{i}}}{\hat{m}_\rmi}}\hat{v}_{\para\rmi}\pp{\hat{g}_{\rmi\bm{k}_\perp}}{\hat{z}} + \sqrt{\f{\hat{T}_{\mr{i}}}{\hat{m}_\rmi}}\hat{v}_{\para\rmi} \pp{}{\hat{z}}\lf( J_0(\hat{a}_\rmi)\f{Z\hat{\phi}_{\bm{k}_\perp}}{\hat{T}_\rmi} + 2\hat{v}_{\perp\rmi}^2\f{J_1(\hat{a}_\rmi)}{\hat{a}_\rmi}\widehat{\delta B}_{||\bm{k}_\perp} \ri) + \f{1}{2}\Big\{ \langle \hat{\chi} \rangle_{\bm{R}_\rmi} ,\hat{h}_\rmi \Big\}_{\bm{k}_\perp} \\
  = -\f{Z}{\sqrt{\hat{m}_\rmi\hat{T}_{\mr{i}}}}\hat{v}_{\para \rmi}J_0(\hat{a}_\rmi)\pp{\hat{A}_{||\bm{k}_\perp}}{\hat{t}} + C_\mr{GK}\lf[\hat{h}_{\rmi\bm{k}_\perp}\ri],
  \label{e:gki normalized}
\end{multline}
\begin{subequations}
\begin{multline}
  \pp{}{\hat{t}}\lf( \f{\widehat{\delta n}_{\rme\bm{k}_\perp}}{\hat{n}_\rme} - \widehat{\delta B}_{||\bm{k}_\perp} \ri) + \f{1}{2}\Bigg\{\hat{\phi} - \f{\hat{T}_\rmi}{\tau}\f{\widehat{\delta n}_\rme}{\hat{n}_\rme},\, \f{\widehat{\delta n}_\rme}{\hat{n}_\rme} - \widehat{\delta B}_{||\bm{k}_\perp}\Bigg\}_{\bm{k}_\perp} + \pp{\hat{u}_{||\rme\bm{k}_\perp}}{\hat{z}}  \\
  = \f{1}{2}\big\{\hat{A}_{||},\, \hat{u}_{||\rme}\big\}_{\bm{k}_\perp} 
  - \hat{\nu}_h \lf( \f{\hat{k}_\perp}{\hat{k}_\mr{max}} \ri)^{2n}\lf( \f{\widehat{\delta n}_{\rme\bm{k}_\perp}}{\hat{n}_{\rme}} - \f{\tau}{\hat{T}_{\rmi}}\hat{\phi}_{\bm{k}_\perp} \ri),
  \label{e:delta ne evo normalized} 
\end{multline}
\begin{equation}
  \pp{\hat{A}_{||\bm{k}_\perp}}{\hat{t}} + \f{1}{2}\lf\{ \hat{\phi} - \f{\hat{T}_\rmi}{\tau}\f{\widehat{\delta n}_\rme}{\hat{n}_\rme},\, \hat{A}_{||} \ri\}_{\bm{k}_\perp} + \pp{\hat{\phi}_{\bm{k}_\perp}}{\hat{z}} - \f{\hat{T}_\rmi}{\tau}\pp{}{\hat{z}}\lf( \f{\widehat{\delta n}_{\rme\bm{k}_\perp}}{\hat{n}_\rme} \ri) = 0,
  \label{e:A|| evo normalized}
\end{equation}
\end{subequations}
\begin{subequations}
\begin{align}
  & \f{\widehat{\delta n}_{\rme\bm{k}_\perp}}{\hat{n}_\rme} = \lf[ \Gamma_0(\hat{\alpha}_\rmi) - 1 \ri]\f{Z\hat{\phi}_{\bm{k}_\perp}}{\hat{T}_\rmi} + \Gamma_1(\hat{\alpha}_\rmi)\delta B_{\para\bm{k}_\perp} + \int\widehat{\rmd^3 \bm{v}}_\rmi\, J_0(\hat{a}_\rmi)\hat{g}_{\rmi\bm{k}_\perp} 
  \label{e:quasineutrality normalized} \\
  & \hat{u}_{\para\rme\bm{k}_\perp} = -\f{\hat{k}_\perp^2\hat{A}_{\para\bm{k}_\perp}}{2\beta_0 Z \hat{n}_\rmi} + \sqrt{\f{\hat{T}_{\mr{i}}}{\hat{m}_\rmi}}\int\widehat{\rmd^3 \bm{v}}_\rmi\, \hat{v}_{\para\rmi} J_0(\hat{a}_\rmi) \hat{g}_{\rmi\bm{k}_\perp},
  \label{e:para Ampere normalized} \\
  & \f{Z}{\tau}\f{\widehat{\delta n}_{\rme\bm{k}_\perp}}{\hat{n}_\rme} + \lf[ \f{2}{\hat{\beta}_0\hat{n}_\rmi\hat{T}_\rmi} + \Gamma_2(\hat{\alpha}_\rmi) \ri]\widehat{\delta B}_{||\bm{k}_\perp} = \lf[ 1 - \Gamma_1(\hat{\alpha}_\rmi) \ri]\f{Z\hat{\phi}_{\bm{k}_\perp}}{\hat{T}_\rmi} - \int\widehat{\rmd^3 \bm{v}}_\rmi\,2\hat{v}_{\perp\rmi}^2\f{J_1(\hat{a}_\rmi)}{\hat{a}_\rmi}\hat{g}_{\rmi\bm{k}_\perp}.
  \label{e:perp Ampere normalized}
\end{align}
\end{subequations}
where
\begin{equation}
  \hat{a}_\rmi = \sqrt{\hat{m}_\rmi\hat{T}_\rmi}\f{\hat{k}_\perp\hat{v}_{\perp\rmi}}{Z}, \quad
  \hat{\alpha}_\rmi = \f{\hat{m}_\rmi\hat{T}_\rmi}{Z^2}\f{\hat{k}_\perp^2}{2}.
\end{equation}
Henceforth, we omit the hat symbols in order to simplify notation.

\section{Numerical Algorithm}\label{s:algorithm}
In this section, we describe the numerical algorithm used to solve the GKI/ITEF equations \eqref{e:gki normalized}, \eqref{e:delta ne evo normalized}--\eqref{e:A|| evo normalized}, and \eqref{e:quasineutrality normalized}--\eqref{e:perp Ampere normalized}.
We start by providing a brief overview of the algorithm used in \agk\ for solving the FGK equations (detailed in Ref.~\cite{Numata2010JCP}) before detailing the modifications we made to solve the GKI/ITEF equations.

\subsection{Implicit time advance for FGK linear terms}\label{ss:FGK implicit algorithm}
In \agk, the linear terms in the gyrokinetic equation~\eqref{e:dgdt} is solved implicitly together with Maxwell's equations \eqref{e:quasineutrality FGK}--\eqref{e:perp Ampere FGK} via a Green's function approach~\cite{Kotschenreuther1995CPC}.
For the linear terms, we employ a Fourier-spectral method in the perpendicular plane, ($x, y$), and a compact finite-difference method to evaluate a derivative in parallel direction, $\p_z$.
The nonlinear terms is treated by pseudo-spectral method (see Section~\ref{ss:nonlinear terms}).
Since there is no explicit perpendicular derivative, $\p_x$ and $\p_y$, except for the nonlinear terms, all the equations are independent with respect to the Fourier modes.
The velocity space is spanned by three variables, pitch angle $\lambda = v_\perp^2/v^2$, energy $E = v_\perp^2 + v_\para^2$, and the sign of the parallel velocity $\sigma = \mr{sgn}(v_\para)$.
Below, we omit the species index $s$ since it is unnecessary here. 
Let us denote the discretized fields by
\begin{equation}
  g_{\rmi\bm{k}_\perp} = g^n_{i,p,q},\quad \f{\delta n_{\rme\bm{k}_\perp}}{n_\rme} = \eta^n_{i},\quad u_{||\rme\bm{k}_\perp} = {u_\para}^n_{i},\quad \phi_{\bm{k}_\perp} = \phi^n_{i},\quad
  A_{||\bm{k}_\perp} = {A_\para}^n_{i},\quad \delta B_{||\bm{k}_\perp} = {B_\para}^n_{i}.
\end{equation}
with the indices corresponding to time grids $t_n = \sum_{j=1}^n \Delta t_j\; (n = 1,\cdots, n_t)$, parallel space grids $z_i = i\Delta z\; (i = 1,\cdots, n_z)$, and velocity space grids $E_p\, (p = 1,\cdots, n_E)$ and $\lambda_q\, (q = 1,\cdots, n_\lambda)$.
The index for $\sigma$ is omitted.
Here $\Delta t_j$ is an adaptive timestep which is modified when the advection speed violates the Courant--Friedrichs--Lewy (CFL) condition~\cite{Courant1967} or considerably larger than the maximum timestep determined by the CFL condition (see Section~\ref{ss:nonlinear terms}).
Allowing temporal and spatial implicitness by parameters $r_t$ and $r_z$, the derivatives with respect to $t$ and $z$ are evaluated by
\begin{equation}
  \pp{f}{t} = \f{1}{2}\lf[ (1 - r_z)\f{f^{n+1}_{i} - f^{n}_{i}}{\Delta t} + (1 + r_z)\f{f^{n+1}_{i+1} - f^{n}_{i+1}}{\Delta t} \ri] \quad \text{and} \quad
  \pp{f}{z} = r_t\f{f^{n}_{i+1} - f^{n}_{i}}{\Delta z} + (1 - r_t)\f{f^{n+1}_{i+1} - f^{n+1}_{i}}{\Delta z},
  \label{e:derivatives}
\end{equation}
where $0 < r_t < 1$ ($r_t = 0$ for fully implicit and $r_t = 1$ for fully explicit) and $0 < r_z < 1$ ($r_z = 0$ for central difference and $r_z = 1$ for first order upwind difference).
Especially when $r_t = 1/2$ and $r_z = 0$, i.e., space and time centered~\cite{Beam1976}, the scheme has second order accuracy both in space and time, and unconditionally stable. 

The discretization of \eqref{e:dgdt} may be written symbolically as, 
\begin{equation}
  a_1 g^{n}_{j} + a_2 g^{n}_{j+1} + b_1 g^{n+1}_{j} + b_2 g^{n+1}_{j+1} = \bm{c}_1\cdot\bm{\Psi}^{n}_j + \bm{c}_2\cdot\bm{\Psi}^{n}_{j+1} + \bm{d}_1\cdot\bm{\Psi}^{n+1}_j + \bm{d}_2\cdot\bm{\Psi}^{n+1}_{j+1} + \calN_{j} + \calN_{j+1},
  \label{e:dgdt d}
\end{equation}
where $\bm{\Psi} = (\phi, A_\para, B_\para)$, the coefficients depend on Fourier space ($\bm{k}_\perp$) and velocity space ($\lambda$ and $E$), $\calN$ represents the nonlinear term, and the velocity space indices $(p, q)$ are omitted. 
Maxwell's equations~\eqref{e:quasineutrality FGK}--\eqref{e:perp Ampere FGK} are compactly written as  
\begin{equation}
  \sfM\cdot\bm{\Psi}^{n+1}_i = \sum_p\sum_q \bm{f}_{p,q}\, g^{n+1}_{i,p,q}, 
  \label{e:Maxwell d}
\end{equation}
where $\sfM$ is a block $3\times3$ matrix with each block as $n_z\times n_z$ submatrix, and the right hand side represents the velocity space integrals appearing in \eqref{e:quasineutrality normalized}--\eqref{e:perp Ampere normalized}.
We note that $\sfM$ and $\bm{f}$ depend neither on $t$ nor $z$.
One may straightforwardly obtain $g^{n+1}$ by substituting~\eqref{e:Maxwell d} into \eqref{e:dgdt d}.
However, such a brute force method is not practical since it requires an inversion of a dense $(n_z n_\lambda n_E n_\sigma n_s)^2$ size matrix.

Kotschenreuther et al. found that the linear property of the equation enables us to break this large matrix into many small matrices~\cite{Kotschenreuther1995CPC}, and by doing so, computational efficiency dramatically improves (see~\cite{Numata2010JCP} for the detailed estimate).
Since \eqref{e:dgdt d} is linear in $g^{n+1}$, its solution is a linear combination of solutions to parts of the equation. 
We split the distribution function at timestep $n+1$ into homogeneous and inhomogeneous parts, viz. $g^{n+1} = g^\mr{(h)} + g^\mr{(inh)}$ and introduce an intermediate timestep variable of EM fields, $\bm{\Psi}^* = \bm{\Psi}^{n+1} - \bm{\Psi}^{n}$.
Then \eqref{e:dgdt d} is split into
\begin{subequations}
\begin{align}
  & a_1 g^{n}_{j} + a_2 g^{n}_{j+1} + b_1 g^\mr{(inh)}_{j} + b_2 g^\mr{(inh)}_{j+1} = \bm{c}_1\cdot\bm{\Psi}^{n}_j + \bm{c}_2\cdot\bm{\Psi}^{n}_{j+1} + \bm{d}_1\cdot\bm{\Psi}^{n}_j + \bm{d}_2\cdot\bm{\Psi}^{n}_{j+1} + \calN_{j} + \calN_{j+1},
  \label{e:dgdt d inhomogeneous} \\
  & b_1 g^\mr{(h)}_{j} + b_2 g^\mr{(h)}_{j+1} = \bm{d}_1\cdot\bm{\Psi}^{*}_j + \bm{d}_2\cdot\bm{\Psi}^{*}_{j+1}.
  \label{e:dgdt d homogeneous}
\end{align}
\end{subequations}
Now, $g^\mr{(inh)}_i$ is immediately obtained by solving \eqref{e:dgdt d inhomogeneous}.
We formally rewrite \eqref{e:dgdt d homogeneous} as
\begin{equation}
  g^{(\mr{h})}_{i,p,q}(\bm{\Psi}^{*}) = \sum_j\lf( \de{g}{\bm{\Psi}} \ri)_{ijpq}\cdot\bm{\Psi}^{*}_j,
  \label{e:dgdt d homogeneous -2-}
\end{equation}
where $(\delta g/\delta\bm{\Psi})_{ijpq}$ is a so-called response matrix which is obtained by the following procedure.
For a given integer $l$, we substitute trial functions $\phi^*_j = \delta_{jl}$, ${A_\para}^*_j = 0$, and ${B_\para}^*_j = 0$ with the Kronecker's delta $\delta_{ij}$ into \eqref{e:dgdt d homogeneous} and solve it for $g^\mr{(h)}_{j,p,q}$.
Then the obtained $g^\mr{(h)}_{j,p,q}$ is equivalent to $(\delta g/\delta \phi)_{jlpq}$.
Recursion of this process over $l = 1,\cdots,n_z$ yields the complete set of $(\delta g/\delta \phi)_{jlpq}$. 
The other components of the response matrix, $(\delta g/\delta A_\para)$ and $(\delta g/B_\para)$, are obtained by using the trial functions in the same way for ${A_\para}^*_j$ and ${B_\para}^*_j$.
Substituting \eqref{e:dgdt d homogeneous -2-} into \eqref{e:Maxwell d}
and moving the terms including $\bm{\Psi}^*$ to the left hand side and the others to the right hand side, we get
\begin{equation}
  \lf[ \sfM\delta_{ij} - \sum_p\sum_q\bm{f}_{p,q}\sum_j\lf( \de{g}{\bm{\Psi}} \ri)_{ijpq} \ri]\cdot\bm{\Psi}^*_j = -\sfM\cdot\bm{\Psi}^n_i + \sum_p\sum_q\bm{f}_{p,q}\, g^\mr{(inh)}_{i,p,q}.
  \label{e:Maxwell d -2-}
\end{equation}
We obtain $\bm{\Psi}^*$ by inversion of the coefficient matrix.
Successively $g^{n+1}$ is obtained by \eqref{e:dgdt d}.
We calculate the coefficient matrix in the initialization step and keep using it unless $\Delta t$ is modified by the CFL condition (see Section~\ref{ss:nonlinear terms}). 

\subsection{Time integration algorithm for the hybrid code}\label{ss:GKI/ITEF algorithm}
Next we consider the time advance algorithm for \eqref{e:gki normalized}, \eqref{e:delta ne evo normalized}--\eqref{e:A|| evo normalized} and \eqref{e:quasineutrality normalized}--\eqref{e:perp Ampere normalized}. 
Using the finite difference \eqref{e:derivatives}, we discretize \eqref{e:delta ne evo normalized} and \eqref{e:A|| evo normalized} as
\begin{multline}
  \f{1}{2}\lf[ (1 - r_z)\f{\eta^{n+1}_{i} - \eta^{n}_{i}}{\Delta t} + (1 + r_z)\f{\eta^{n+1}_{i+1} - \eta^{n}_{i+1}}{\Delta t} \ri]
  - \f{1}{2}\lf[ (1 - r_z)\f{{B_\para}^{n+1}_{i} - {B_\para}^{n}_{i}}{\Delta t} + (1 + r_z)\f{{B_\para}^{n+1}_{i+1} - {B_\para}^{n}_{i+1}}{\Delta t} \ri] \\
  + r_t\f{{u_\para}^{n}_{i+1} - {u_\para}^{n}_{i}}{\Delta z} + (1 - r_t)\f{{u_\para}^{n+1}_{i+1} - {u_\para}^{n+1}_{i}}{\Delta z}
  =
  - \f{\nu_h}{4}\lf( \f{k_\perp}{k_\mr{max}} \ri)^{2n}\Bigg\{ \lf[ (1 - r_z)\lf( \eta^{n+1}_{i} + \eta^{n}_{i} \ri) + (1 + r_z)\lf( \eta^{n+1}_{i+1} + \eta^{n}_{i+1} \ri) \ri] \\
  - \f{\tau}{T_{0\rmi}}\lf[ (1 - r_z)\lf( \phi^{n+1}_{i} + \phi^{n}_{i} \ri) + (1 + r_z)\lf( \phi^{n+1}_{i+1} + \phi^{n}_{i+1} \ri) \ri] \Bigg\} 
  + \f{1}{2}\lf[ (1 + r_z)\calN^{(\eta)}_{i+1} + (1 - r_z)\calN^{(\eta)}_{i} \ri]
  \label{e:delta ne evo normalized d}
\end{multline}
\begin{multline}
  \f{1}{2}\lf[ (1 - r_z)\f{{A_\para}^{n+1}_{i} - {A_\para}^{n}_{i}}{\Delta t} + (1 + r_z)\f{{A_\para}^{n+1}_{i+1} - {A_\para}^{n}_{i+1}}{\Delta t} \ri]
  + r_t\f{\phi^{n}_{i+1} - \phi^{n}_{i}}{\Delta z} + (1 - r_t)\f{\phi^{n+1}_{i+1} - \phi^{n+1}_{i}}{\Delta z} \\
  =
  \f{T_\rmi}{\tau}\lf[r_t\f{\eta^{n}_{i+1} - \eta^{n}_{i}}{\Delta z} + (1 - r_t)\f{\eta^{n+1}_{i+1} - \eta^{n+1}_{i}}{\Delta z}\ri] + \f{1}{2}\lf[ (1 + r_z)\calN^{(A_\para)}_{i+1} + (1 - r_z)\calN^{(A_\para)}_{i} \ri]
  \label{e:A|| evo normalized d}
\end{multline}
We may choose the value of $r_t$ and $r_z$ independently from the ion gyrokinetic equation \eqref{e:gki normalized}.
The Maxwell's equations \eqref{e:quasineutrality normalized}--\eqref{e:perp Ampere normalized} are discretized as
\begin{subequations}
\begin{align}
  & Zn_\rmi\eta^{n+1}_i = -\pzcG_0\phi^{n+1}_i + \pzcG_1 {B_\para}^{n+1}_{i} + \calM^{(0)}\lf( g^{n+1}_i \ri)
  \label{e:quasineutrality normalized d} \\
  & 2\beta_0 Z n_\rmi{u_\para}^{n+1}_i = -k_\perp^2 {A_\para}^{n+1}_i + \calM^{(1)}\lf( g^{n+1}_i \ri)
  \label{e:para Ampere normalized d} \\
  & \f{n_\rmi T_\rmi Z}{2\tau}{\eta}^{n+1}_i + \lf( \f{1}{\beta_0} + \pzcG_2 \ri){B_\para}^{n+1}_i = \f{1}{2}\lf( n_\rmi Z - \pzcG_1 \ri) {\phi}^{n+1}_i - \calM^{(2)}\lf( g^{n+1}_i \ri)
  \label{e:perp Ampere normalized d}
\end{align}
\end{subequations}
Here the coefficients are defined by
\begin{equation}
  \pzcG_0 = \f{Z^2 n_\rmi}{T_\rmi}\lf[ 1 - \Gamma_0(\alpha_\rmi) \ri],\quad
  \pzcG_1 = Zn_\rmi\Gamma_1(\alpha_\rmi),\quad
  \pzcG_2 = \f{1}{2}n_\rmi T_\rmi\Gamma_2(\alpha_\rmi),
\end{equation}
the nonlinear terms are represented by
\begin{subequations}
\begin{align}
  & \calN^{(A_\para)} = -\f{1}{2}\Big\{ \phi - \f{T_\rmi}{\tau}\eta,\, A_{||} \Big\}\\
  & \calN^{(\eta)} = -\f{1}{2}\Big\{\phi - \f{T_\rmi}{\tau}\eta,\, \eta - \delta B_{||\bm{k}_\perp}\Big\} + \f{1}{2}\Big\{A_{||},\, u_{||\rme}\Big\},
\end{align}
\end{subequations}
and the velocity integral operators are defined by
\begin{subequations}
\begin{align}
  \calM^{(0)}\lf( g_{\rmi\bm{k}_\perp} \ri) = n_\rmi Z\int\rmd^3 \bm{v}_\rmi\, J_0(a_\rmi)g_{\rmi\bm{k}_\perp}\\
  \nonumber\\
  \calM^{(1)}\lf( g_{\rmi\bm{k}_\perp} \ri) = 2\beta_0 n_\rmi Z\sqrt{\f{T_{\mr{i}}}{m_\rmi}}\int\rmd^3 \bm{v}_\rmi\, v_{\para\rmi} J_0(a_\rmi) g_{\rmi\bm{k}_\perp}\\
  \nonumber\\
  \calM^{(2)}\lf( g_{\rmi\bm{k}_\perp} \ri) = n_\rmi T_\rmi\int\rmd^3 \bm{v}_\rmi\, v_{\perp \rmi}^2\f{J_1(a_\rmi)}{a_\rmi} g_{\rmi\bm{k}_\perp}.
\end{align}
\end{subequations}
Compared to FGK equations (\eqref{e:dgdt d} and \eqref{e:Maxwell d}), the electron's gyrokinetic equation is replaced by \eqref{e:delta ne evo normalized d} and \eqref{e:A|| evo normalized d}; hence, in principle, it is possible to separate $\eta^{n+1}$ and $u_\para^{n+1}$ into homogeneous and inhomogeneous parts in the same way as the \agk\ algorithm.
However, here we employ a more straightforward way; 
$\eta^{n+1}$ and $u_\para^{n+1}$ are simply eliminated from \eqref{e:delta ne evo normalized d}, \eqref{e:A|| evo normalized d}, and \eqref{e:perp Ampere normalized d} by using \eqref{e:quasineutrality normalized d} and \eqref{e:para Ampere normalized d}.
The resulting equations are compactly written as
\begin{subequations}
\begin{align}
  & \sfA_{ij}\phi^{n+1}_j + \sfB_{ij}{A_\para}^{n+1}_j + \sfC_{ij}{B_\para}^{n+1}_j - \sfD_{ij}\lf( {B_\para}^{n+1}_j - {B_\para}^n_j \ri) = (\sfD_{ij} + \sfE_{ij})\eta^n_j - \sfF_{ij}{u_\para}^n_j + \f{\tau}{T_\rmi}\sfE_{ij}\phi^n_j \nonumber\\
  & \hspace{7em} + \sfG_{ij}\calM^{(0)}\lf( g^{n+1}_j \ri) + \sfH_{ij}\calM^{(1)}\lf( g^{n+1}_j \ri) + \f{1}{2}\lf[ (1 + r_z)\calN^{(\eta)}_{i+1} + (1 - r_z)\calN^{(\eta)}_{i} \ri],
  \label{e:delta ne evo normalized d -2-} \\
  & \sfI_{ij}\phi^{n+1}_j + \sfF_{ij}\phi^{n}_j + \sfD_{ij}\lf( {A_\para}^{n+1}_j - {A_\para}^{n}_j \ri) + \sfJ_{ij}{B_\para}^{n+1}_j = \nonumber\\
  & \hspace{5em} \sfK_{ij}\eta^n_j + \sfL_{ij}\calM^{(0)}\lf( g^{n+1}_j \ri) + \f{1}{2}\lf[ (1 + r_z)\calN^{(A_\para)}_{i+1} + (1 - r_z)\calN^{(A_\para)}_{i} \ri]
  \label{e:A|| evo normalized d -2-}, \\
  & a\phi^{n+1}_i + b{B_\para}_i =  -\f{T_\rmi}{2\tau}\calM^{(0)}\lf( g^{n+1}_i \ri) - \calM^{(2)}\lf( g^{n+1}_i \ri),
  \label{e:perp Ampere normalized d -2-} 
\end{align}
\end{subequations}
where
\begin{align}
  &
  \sfA = -\lf[ \f{1}{2\Delta t}\f{\pzcG_0}{Z n_\rmi} + \f{\nu_h}{4}\lf( \f{k_\perp}{k_\mr{max}} \ri)^{2n}\lf( \f{\pzcG_0}{Zn_{\rmi}} + \f{\tau}{T_{\rmi}} \ri) \ri]\sfT, \quad
  \sfB = -\f{1 - r_t}{\Delta z}\f{k_\perp^2}{2\beta_0Zn_\rmi}\sfZ, \quad
  \sfC = \lf[ \f{1}{2\Delta t}\f{\pzcG_1}{Z n_\rmi} + \f{\nu_h}{4}\lf( \f{k_\perp}{k_\mr{max}} \ri)^{2n}\f{\pzcG_1}{Zn_{\rmi}} \ri]\sfT, \nonumber\\
  &
  \sfD = \f{1}{2\Delta t}\sfT, \quad
  \sfE = -\f{\nu_h}{4}\lf( \f{k_\perp}{k_\mr{max}} \ri)^{2n}\sfT, \quad\
  \sfF = \f{r_t}{\Delta z}\sfZ, \quad
  \sfG = -\lf[ \f{1}{2\Delta t} + \f{\nu_h}{4}\lf( \f{k_\perp}{k_\mr{max}} \ri)^{2n} \ri]\f{1}{Z n_\rmi}\sfT, \quad
  \sfH = -\f{1 - r_t}{\Delta z}\f{1}{2\beta_0Zn_\rmi}\sfZ, \nonumber\\
  &
  \sfI = \f{1 - r_t}{\Delta z}\lf( 1 + \f{T_\rmi}{Z \tau n_\rmi} \pzcG_0\ri)\sfZ, \quad\
  \sfJ = -\f{1 - r_t}{\Delta z}\f{T_\rmi}{Z \tau n_\rmi} \pzcG_1\sfZ, \quad
  \sfK = \f{r_t}{\Delta z}\f{T_\rmi}{\tau}\sfZ, \quad
  \sfL = \f{1 - r_t}{\Delta z}\f{T_\rmi}{Z \tau n_\rmi} \sfZ, \nonumber\\
  \nonumber\\
  &
  \sfZ = 
  \lf( 
    \begin{array}{ccccc}
      -1 &  1 &        &    &    \\
         & -1 &        &    &    \\
         &    & \ddots &    &    \\
         &    &        & -1 &  1 \\
       1 &    &        &    & -1 \\
    \end{array} 
  \ri), \quad
  \sfT = 
  \lf( 
    \begin{array}{ccccc}
      1 - r_z & 1 + r_z &        &         &         \\
              & 1 - r_z &        &         &         \\
              &         & \ddots &         &         \\
              &         &        & 1 - r_z & 1 + r_z \\
      1 + r_z &         &        &         & 1 - r_z \\
    \end{array} 
  \ri) \nonumber\\
  &
  a = -\f{1}{2}\lf( \f{T_\rmi}{\tau}\pzcG_0 + n_\rmi Z - \pzcG_1 \ri), \quad
  b = \lf( \f{T_\rmi}{2\tau}\pzcG_1 + \f{1}{\beta_0} + \pzcG_2 \ri),
\end{align}
and the repeated indices $j$ is summed over $j = 1, \cdots, n_z - 1$.
Here the sparse matrices $\sfZ$ and $\sfT$ originate from the derivatives in $z$ and $t$, respectively, and the left bottom corner slots are contributions from the periodic boundary condition for $z$. 
These three equations are corresponding to \eqref{e:Maxwell d} in the \agk\ algorithm.
The important difference between \eqref{e:delta ne evo normalized d -2-}--\eqref{e:perp Ampere normalized d -2-} and \eqref{e:Maxwell d} is that \eqref{e:delta ne evo normalized d -2-} and \eqref{e:A|| evo normalized d -2-} have communications with the values on different $z_i$ whereas \eqref{e:Maxwell d} does not.
This communication originates from the derivatives with respect to $t$ and $z$ in \eqref{e:delta ne evo normalized} and \eqref{e:A|| evo normalized}.

Next we introduce the intermediate value $\bm{\Psi}^*$ and split the ion distribution function into the homogeneous and inhomogeneous parts in the same manner as the \agk\ algorithm.
Moving the intermediate value to the left hand side and the rest to the right hand side, \eqref{e:delta ne evo normalized d -2-}--\eqref{e:perp Ampere normalized d -2-} are summarized in
\begin{equation}
  \lf( 
    \begin{array}{ccc}
      \sfP_{11} & \sfP_{12} & \sfP_{13} \\
      \sfP_{21} & \sfP_{22} & \sfP_{23} \\
      \sfP_{31} & \sfP_{32} & \sfP_{33} \\
    \end{array} 
  \ri)
  \lf( 
    \begin{array}{c}
      \phi^*_k \\
      {A_\para}^*_k \\
      {B_\para}^*_k \\
    \end{array} 
  \ri) 
  =
  \lf( 
    \begin{array}{c}
      \sfQ_1 \\
      \sfQ_2 \\
      \sfQ_3 \\
    \end{array} 
  \ri), 
  \label{e:Maxwell-ITEF} 
\end{equation}
with
\begin{subequations}
\begin{align}
  \sfP_{11} =\;& \sfA_{ij}\delta_{jk} - \sfG_{ij}\calM^{(0)}\lf( \lf( \de{g}{\phi} \ri)_{jk} \ri) - \sfH_{ij}\calM^{(1)}\lf( \lf( \de{g}{\phi} \ri)_{jk} \ri), \\
  \sfP_{12} =\;& \sfB_{ij}\delta_{jk} - \sfG_{ij}\calM^{(0)}\lf( \lf( \de{g}{A_\para} \ri)_{jk} \ri) - \sfH_{ij}\calM^{(1)}\lf( \lf( \de{g}{A_\para} \ri)_{jk} \ri), \\
  \sfP_{13} =\;& \lf( \sfC_{ij} - \sfD_{ij} \ri)\delta_{jk} - \sfG_{ij}\calM^{(0)}\lf( \lf( \de{g}{B_\para} \ri)_{jk} \ri) - \sfH_{ij}\calM^{(1)}\lf( \lf( \de{g}{B_\para} \ri)_{jk} \ri), \\
  \sfP_{21} =\;& \sfI_{ij}\delta_{jk} - \sfL_{ij}\calM^{(0)}\lf( \lf( \de{g}{\phi} \ri)_{jk} \ri), \\
  \sfP_{22} =\;& \sfD_{ij}\delta_{jk} - \sfL_{ij}\calM^{(0)}\lf( \lf( \de{g}{A_\para} \ri)_{jk} \ri), \\
  \sfP_{23} =\;& \sfJ_{ij}\delta_{jk} - \sfL_{ij}\calM^{(0)}\lf( \lf( \de{g}{B_\para} \ri)_{jk} \ri), \\
  \sfP_{31} =\;& a\delta_{ik} + \f{T_\rmi}{2\tau}\calM^{(0)}\lf( \lf( \de{g}{\phi} \ri)_{ik} \ri) + \calM^{(2)}\lf( \lf( \de{g}{\phi} \ri)_{ik} \ri), \\
  \sfP_{32} =\;& 0, \\
  \sfP_{33} =\;& b\delta_{ik} + \f{T_\rmi}{2\tau}\calM^{(0)}\lf( \lf( \de{g}{B_\para} \ri)_{ik} \ri) + \calM^{(2)}\lf( \lf( \de{g}{B_\para} \ri)_{ik} \ri), \\
  \sfQ_{1} =\;& (\sfD_{ij} + \sfE_{ij})\eta^n_j - \sfF_{ij}{u_\para}^n_j - \lf( \sfA_{ij} - \f{\tau}{T_\rmi}\sfE_{ij} \ri)\phi^{n}_j - \sfB_{ij}{A_\para}^{n}_j - \sfC_{ij}{B_\para}^{n+1}_j \nonumber\\
  & + \sfG_{ij}\calM^{(0)}\lf( g^\mr{(inh)}_j \ri) + \sfH_{ij}\calM^{(1)}\lf( g^\mr{(inh)}_j \ri) + \f{1}{2}\lf[ (1 + r_z)\calN^{(\eta)}_{i+1} + (1 - r_z)\calN^{(\eta)}_{i} \ri]  \\
  \sfQ_{2} =\;& \sfK_{ij}\eta^n_j - \lf( \sfI_{ij} + \sfF_{ij} \ri)\phi^{n}_j - \sfJ_{ij}{B_\para}^{n+1}_j + \sfL_{ij}\calM^{(0)}\lf( g^\mr{(inh)}_j \ri) + \f{1}{2}\lf[ (1 + r_z)\calN^{(A_\para)}_{i+1} + (1 - r_z)\calN^{(A_\para)}_{i} \ri]  \\
  \sfQ_{3} =\;& -a\phi^{n}_i - b{B_\para}_i - \f{T_\rmi}{2\tau}\calM^{(0)}\lf( g^\mr{(inh)}_i \ri) - \calM^{(2)}\lf( g^\mr{(inh)}_i \ri),
\end{align}
\end{subequations}
where the velocity space indices $(p, q)$ for $(\delta g/\delta\bm{\Psi})$ do not appear because of the velocity space integral.
The obtained equation \eqref{e:Maxwell-ITEF} is a ``compound'' of ITEF and Maxwell's equations.
This equation corresponds to \eqref{e:Maxwell d -2-} in the \agk\ algorithm.
Although the matrix $\sfP$ is more complicated than the coefficient matrix in \eqref{e:Maxwell d -2-}, both matrices are dense with the same size, resulting in the same computational cost for the inversion.
In the same way for \agk, we calculate $\sfP$ in the initialization step and keep using it for the rest of the computation unless $\Delta t$ is adjusted (see Section~\ref{ss:nonlinear terms}).
By inverting $\sfP$, $\bm{\Psi}^*$ is obtained, and subsequently, the ion distribution function $g^{n+1}$ is calculated. 
Finally, $\eta^{n+1}$ and $u_\para^{n+1}$ are calculated by plugging $\bm{\Psi}^{n+1}$ and $g^{n+1}$ into \eqref{e:quasineutrality normalized d} and \eqref{e:para Ampere normalized d}.
Thus all the variables at $n+1$ timestep is obtained.

\subsection{Explicit treatment of nonlinear terms}\label{ss:nonlinear terms}
The nonlinear terms in \eqref{e:gki normalized} and \eqref{e:delta ne evo normalized}--\eqref{e:A|| evo normalized} are calculated by pseudo-spectral method, i.e., evaluated in the real space then Fourier transformed to the wave number space.
The 2/3 truncation rule is applied for dealiasing~\cite{Orszag1971}.
The calculated nonlinear terms are added to \eqref{e:dgdt d inhomogeneous} by the third order Adams--Bashforth method with variable timestep: 
\begin{equation}
  \calN = c_1 \calN^n + c_2 \calN^{n-1} + c_3 \calN^{n-2}, \\
\end{equation}
with
\begin{subequations}
\begin{align}
  c_1 =\;& 1 + \f{\Delta t_n}{2\Delta t_{n-2}}\lf( \f{\Delta t_{n-1} + \Delta t_{n-2}}{\Delta t_{n-1}} - \f{\Delta t_{n-1}}{\Delta t_{n-1} + \Delta t_{n-2}} \ri) + \f{\Delta t_n^2}{3\Delta t_{n-1}(\Delta t_{n-1} + \Delta t_{n-2})}, \\
  c_2 =\;& -\f{\Delta t_n}{\Delta t_{n-1}\Delta t_{n-2}}\lf(\f{\Delta t_{n-1} + \Delta t_{n-2}}{2} + \f{\Delta t_n}{3}\ri), \\
  c_3 =\;& \f{\Delta t_n}{(\Delta t_{n-1} + \Delta t_{n-2})\Delta t_{n-2}}\lf(\f{\Delta t_{n-1}}{2} + \f{\Delta t_n}{3}\ri),
\end{align}
\end{subequations}
where $\Delta t_n$ corresponds to the timestep at $n$th step.

Since the nonlinear terms are treated by the explicit method, the timestep is restricted by the CFL condition for nonlinear runs.
In order to ensure stable time integration, the adaptive timestep used in \agk~\cite{Numata2010JCP} is also enabled; at each timestep, the CFL condition is evaluated by advection velocity (Section~\ref{s:CFL improvement}), and the next time step is modified when the CFL condition is violated. 

\subsection{External antenna driving}\label{s:antenna}
In the realistic physical problem settings, the system is usually driven by external energy source injected at much larger scale than ion kinetic scale.
In \agk, such an external driving is modeled by an oscillating antenna that excites parallel vector potential $A_{\para\rma}$~\cite{TenBarge2014}.
The gyrokinetic equation \eqref{e:dgdt} and the parallel Ampere's law \eqref{e:para Ampere FGK} are solved with replacing $A_\para$ to $A_\para + A_{\para\rma}$.
This is equivalent to adding external electric field $\bm{E}_\rma = -c^{-1}\p_t A_{\para\rma}\zhat$ to \eqref{e:dgdt} and external current $\bm{J}_\rma = -(ck_\perp^2/4\pi)A_{\para\rma}\zhat$ to \eqref{e:para Ampere FGK}.
This $\bm{J}_\rma$ corresponds to the one in the energy balance equation \eqref{e:FGK energy}.

The antenna driving in \agk\ is also inherited in the present code.
In the same way for GKI/ITEF, we replace $A_\para$ in \eqref{e:gki normalized}, \eqref{e:delta ne evo normalized}--\eqref{e:A|| evo normalized}, and \eqref{e:quasineutrality normalized}--\eqref{e:perp Ampere normalized} by $A_\para + A_{\para\rma}$.
Users may choose the amplitude, number of modes, wave numbers, frequency, and decorrelation rate of the antenna.

\section{Improvement of the computational time}\label{s:CFL improvement}
In this section, we compare the computational cost of the present code to that of \agk.
The greatest concern for nonlinear runs is the restriction of the timestep due to the CFL condition which is determined by the perpendicular advection speed.
The parallel streaming term does not restrict the CFL condition as it is treated implicitly.
Since we eliminate the fast electron motion in the hybrid code, the CFL condition should be alleviated.
Below we estimate the magnitude of the advection speed for FGK and GKI/ITEF equations term-by-term.

We start with the FGK equation \eqref{e:dgdt}.
The perpendicular advection velocity for each species is given by $\bm{v}_{\chi s} = -(c/B_0)\p \lf< \chi \ri>_{\bm{R}_s}/\p \bm{R}_s\times\zhat$.
As shown below, the CFL condition is determined by the electron motion.
Since we focus on the ion kinetic scale where $k_\perp\rho_\rmi \sim 1 \ll k_\perp\rho_\rme = (Z/\sqrt{\tau})\sqrt{m_\rme/m_\rmi}\,k_\perp\rho_\rmi$,  we may assume $a_\rme \ll 1$ in the electron gyrokinetic equation \eqref{e:dgdt}.
A critical balance conjecture~\cite{Goldreich1995}, $k_\para v_\rmA \sim k_\perp u_\perp$ with Alfv\'{e}n speed $v_\rmA = v_\mr{thi}/\sqrt{\beta_\rmi}$ and $E\times B$ velocity $\bm{u}_\perp = -(c/B_0)\nbl\phi\times\zhat$, yields the estimates shown in Ref.~\cite{Schekochihin2009ApJ}:
\begin{subequations}
\begin{align}
  \f{v_\para A_\para}{c} \sim\;& \sqrt{\f{\beta_\rmi}{\tau}}\sqrt{\f{m_\rmi}{m_\rme}}\phi \\
  \f{T_\rme}{e}\f{v_\perp^2}{v_\mr{the}^2}\f{\delta B_\para}{B_0} \sim\;& \f{Z}{\tau}k_\perp\rho_\rmi\sqrt{\beta_\rmi}\phi.
\end{align}
\end{subequations}
By using these, we estimate the advection speed for the electron gyrokinetic equation as
\begin{equation}
  |\bm{v}_{\chi \rme}| \sim \max\lf( 1,\, \sqrt{\f{\beta_\rmi}{\tau}}\sqrt{\f{m_\rmi}{m_\rme}}, \, \f{Z}{\tau}k_\perp\rho_\rmi\sqrt{\beta_\rmi}\ri)u_\perp.
  \label{e:GKE uperp estimate}
\end{equation}
In most astrophysical systems, $\beta_\rmi \gtrsim 1$ and $\tau \gtrsim 1$.
Therefore, the advection speed is of order $\sqrt{(\beta_\rmi/\tau)(m_\rmi/m_\rme)}\,u_\perp$. 

Next, we estimate the advection speed for the GKI/ITEF system.
In the same way as \eqref{e:GKE uperp estimate}, the advection speed for the ion gyrokinetic equation is estimated as
\begin{equation}
  |\bm{v}_{\chi \rmi}| \sim \max\lf( 1,\, \sqrt{\beta_\rmi}, \, k_\perp\rho_\rmi\sqrt{\beta_\rmi}\ri)u_\perp.
  \label{e:GKI uperp estimate}
\end{equation}
In the ITEF equations \eqref{e:delta ne evo} and \eqref{e:A|| evo}, the nonlinear terms, except for the right hand side in \eqref{e:delta ne evo}, are convective derivatives with velocity
\begin{equation}
  -\f{c}{B_0}\nbl_\perp\lf( \phi - \f{T_\rme}{e}\f{\delta n_\rme}{n_\rme} \ri)\times\zhat \sim \max\lf( 1, \f{Z}{\tau}k_\perp\rho_\rmi\sqrt{\beta_\rmi} \ri)u_\perp
  \label{e:A|| evo uperp estimate},
\end{equation}
Therefore, the advection speed for the GKI/ITEF equations is of order $\sqrt{\beta_\rmi}u_\perp$, and thus the maximum timestep determined by the CFL condition should be $\sqrt{m_\rmi/m_\rme} \sim 43$ times greater than FGK.
Furthermore, the size of the array for $h_s$ in the present hybrid code is half that of \agk\ with a single ion species and kinetic electrons. 
In total, the hybrid code runs $2\sqrt{m_\rmi/m_\rme} \sim 100$ times faster than \agk, which makes parameter scans in $\tau$ and $\beta_\rmi$ feasible.

The nonlinear term on the right hand side of \eqref{e:delta ne evo} is not the convective derivative; hence we may not evaluate the convective speed of this term.
However, we may assume that this term does not affect a stable time evolution for the following reason.
We consider this nonlinear term as a source term of \eqref{e:delta ne evo}. 
Substituting \eqref{e:para Ampere}, this term splits into two Poisson brackets which are proportional to $\{ A_\para,\, \nbl_\perp^2 A_\para \}$ and $\{ A_\para,\, \lf< h_\rmi \ri>_\bm{r} \}$.
These terms do not restrict the timestep because the CFL conditions for the ion gyrokinetic equation and $A_\para$ evolution equation \eqref{e:A|| evo} are satisfied.
Moreover, the order of the amplitude of this source term is the same as the convective derivative term:
\begin{equation}
  \f{1}{B_0}\lf\{ A_\para, u_{\para\rme} \ri\} \sim \sqrt{\f{\beta_\rmi}{\tau}}k_\perp u_\perp\epsilon.
\end{equation}
Thus, this nonlinear term does not break the stable time evolution.

\section{Numerical tests}\label{s:tests}
In this section, we present numerical tests demonstrating the validity of the hybrid code.
We first present a linear test to verify the implicit time integral algorithm used for the ITEF (Section~\ref{ss:GKI/ITEF algorithm}), followed by a nonlinear test to show that the nonlinear terms are properly treated by the pseudo-spectral method.
The nonlinear test is conducted for two distinctive spatial scales, viz., MHD inertial range and ion kinetic range.
For both cases, the result of the hybrid code is compared to \agk.

\subsection{Linear decoupled ITEF}\label{ss:decoupled electron fluid}
\begin{figure}[htpb]
  \begin{center}
    \includegraphics*[width=1.0\textwidth]{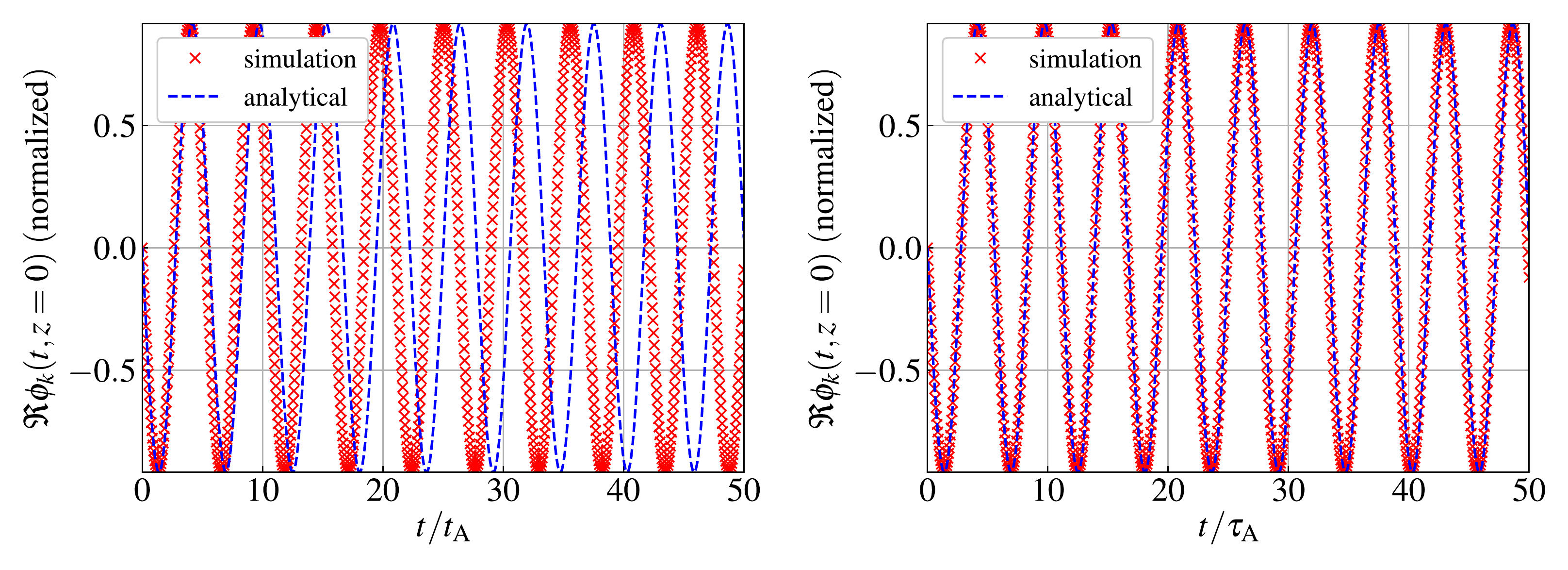}
  \end{center}
  \caption{Time evolution of the real part of $\phi_{\bm{k}_\perp}$ for the one-dimensional linear decoupled ITEF. The number of grids in $z$ direction is set to (\textit{left}) $n_z = 8$ and (\textit{right}) $n_z = 32$.}
  \label{f:decoupled ITEF}
\end{figure}

When the velocity integral terms in Maxwell's equations \eqref{e:quasineutrality normalized}--\eqref{e:perp Ampere normalized} are artificially set to zero, the ITEF equations \eqref{e:delta ne evo normalized} and \eqref{e:A|| evo normalized} are decoupled from the ion gyrokinetic equation.
When the nonlinear terms are also neglected, \eqref{e:delta ne evo normalized}--\eqref{e:perp Ampere normalized} are reduced to a one-dimensional harmonic oscillator equation with normalized frequency 
\begin{equation}
  \omega =\sqrt{\f{k_\perp^2 \lf(\alpha T_\rmi/\tau - (\pzcG_1-\alpha n_\rmi Z)/\pzcG_0\ri)}{2 (\alpha -1) \beta_0  n_\rmi Z}}, \quad \text{where} \quad
  \alpha = \f{(Z n_\rmi - \pzcG_1) \pzcG_1 - 2 (1/\beta_0 + \pzcG_2) \pzcG_0}{Z n_\rmi [Z n_\rmi - \pzcG_1 + (T_\rmi/\tau) \pzcG_0]}.
\end{equation}
In order to verify the implicit time integration scheme described in~Section~\ref{ss:GKI/ITEF algorithm}, we solve this linear decoupled ITEF.
Figure~\ref{f:decoupled ITEF} shows the numerical and analytic solutions with $Z = 1$, $\beta_\rmi = 1$, $\tau = 1$, and $k_\perp \rho_\rmi = 1.0$ for different $n_z$.
We find that the simulation reproduces the analytic solution over many Alfv\'{e}n times, $\tau_\rmA = L_\para/v_\rmA$ with Alfv\'{e}n speed $v_\rmA = v_\mr{thi}/\sqrt{\beta_\rmi}$, if $n_z$ is sufficiently large. 

\subsection{Linear Alfv\'{e}n wave}\label{ss:linear Alfven wave}
\begin{figure}[htpb]
  \begin{center}
    \includegraphics*[width=0.6\textwidth]{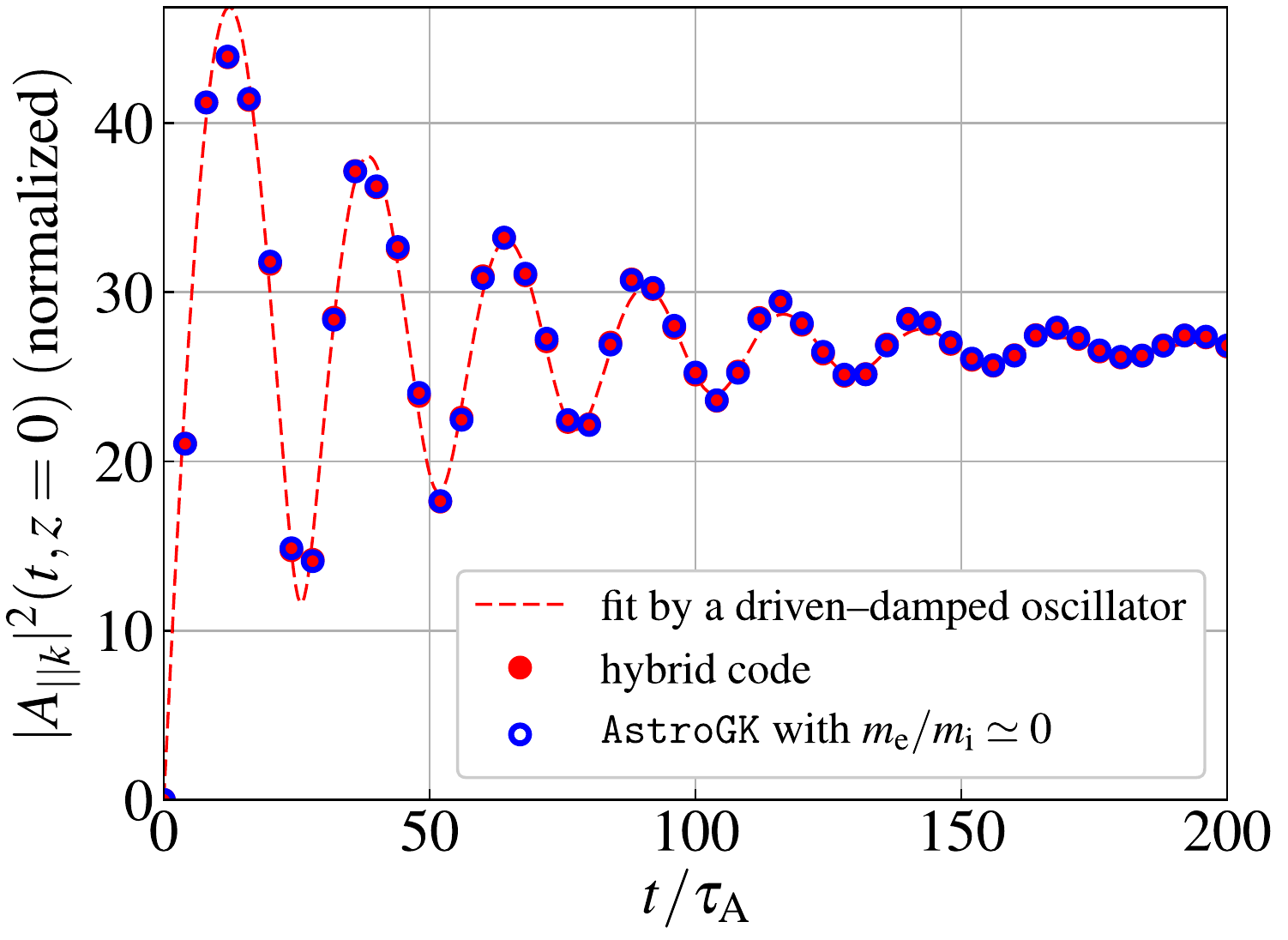}
  \end{center}
  \caption{Time evolution of $A_\para$ solved by the hybrid code (\textit{red closed circle}) and \agk\ with $m_\rme/m_\rmi = 10^{-10}$ (\textit{blue open circle}). The red broken line is a fit of the hybrid code result by an analytical solution of the driven--damped oscillator equation.}
  \label{f:driven damped}
\end{figure}

We next demonstrate that the code correctly captures Alv\'{e}n wave behavior in the appropriate limit, i.e., $m_\rme/m_\rmi \to 0$.
When the velocity integral terms in \eqref{e:quasineutrality normalized}--\eqref{e:perp Ampere normalized} are finite, the linearized GKI/ITEF behaves as a damped oscillator due to the ion Landau damping.
The dispersion relation for the GKI/ITEF is given by setting $m_\rme/m_\rmi = 0$ for the FGK dispersion relation shown in Ref.~\cite{Howes2006ApJ}. 
We compare the time evolution solved by the hybrid code with \agk\ with $m_\rme/m_\rmi = 10^{-10}$ where the Alfv\'{e}n wave is excited by the antenna. 
The plasma parameters are set to $Z = 1$, $\beta_i = 1$, $\tau = 1$, and $k_\perp \rho_\rmi = 1.0$. 
For these parameters, GKI/ITEF has frequency and damping rate for the Alfv\'{e}n wave as $\omega/k_\para v_\rmA = 1.137 - 0.020i$.
The antenna frequency is chosen to be slightly off-resonant, $\omega_a/k_\para v_\rmA = 0.9$. 
The number of grids is set to $(n_z,\, n_\lambda,\, n_E) = (32,\, 8,\, 32)$.
All the fields are initially set to zero. 
Figure~\ref{f:driven damped} shows the time evolution of $A_\para$.
One finds that the results obtained by both codes agree.
The time evolution of GKI/ITEF is fitted by the damped--oscillator solution, then the frequency and damping rate are determined as $\omega/k_\para v_\rmA = 1.142 - 0.022i$ which is close to the analytical value.

\begin{figure}[htpb]
  \begin{center}
    \includegraphics*[width=0.7\textwidth]{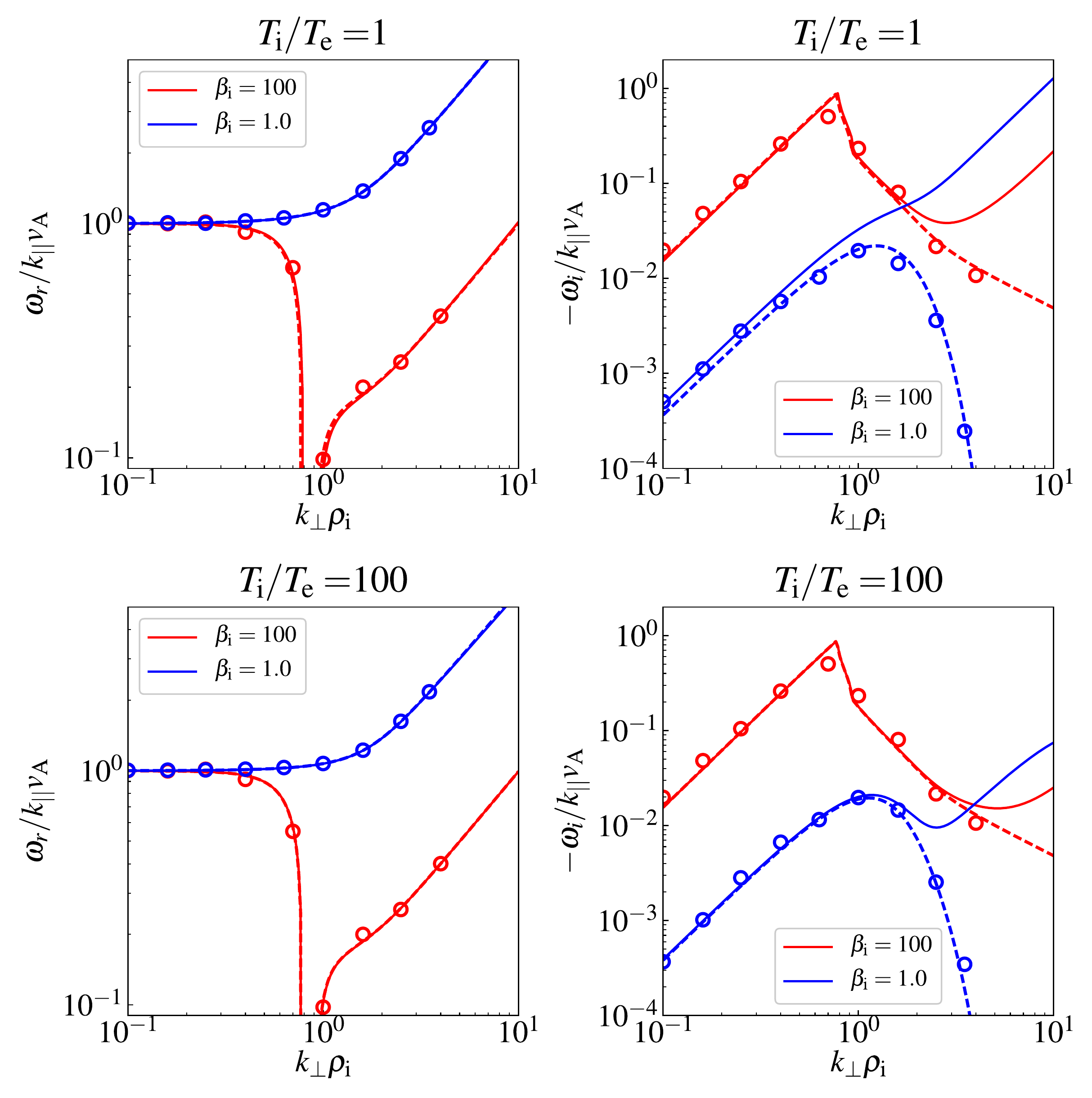}
  \end{center}
  \caption{Frequency (\textit{left}) and damping rate (\textit{right}) of the KAW for $\beta = 1,\, 100$ and $\tau = 1,\, 100$. The solid and broken lines indicate the analytic solutions for FGK and GKI/ITEF, respectively, while the open circles show the result from the simulation.}
  \label{f:KAW dispersion}
\end{figure}

By calculating the time evolution for various $k_\perp \rho_\rmi$, we may construct a dispersion diagram.
Figure~\ref{f:KAW dispersion} shows the numerical and analytic solutions of the frequency and damping rate of the KAW for several parameter cases, $\beta = 1,\, 100$ and $\tau = 1,\, 100$.
The grids used are the same as for the above case.
The results show good agreement between the numerical and analytical solutions.
Whereas the frequencies for FGK and GKI/ITEF are in agreement for all parameters and all $k_\perp\rho_\rmi$, there are appreciable differences in the damping rates between FGK and GKI/ITEF for $k_\perp \rho_\rmi \gtrsim 1$, except for the $\beta_\rmi = 100$ and $\tau = 100$ case.
This discrepancy is due to the missing electron Landau damping.

\subsection{Two dimensional Orszag-Tang problem}\label{ss:2D OT}
In this section, we show results of a nonlinear electromagnetic test known as the Orszag-Tang vortex problem~\cite{Orszag1979}, which has been regularly used to study decaying MHD turbulence.
We compare the results obtained by FGK and GKI/ITEF simulations under the same simulation settings.
While there are several variants of initial conditions for the Orszag-Tang problem~\cite{Biskamp1989,Loureiro2016}, we use an asymmetric initial condition similar to the one proposed in Ref~\cite{Loureiro2016},
\begin{eqnarray}
  \phi(x, y) &=& -\f{B_0}{c}\delta u_0\lf( \f{L_\perp}{2\pi} \ri) \lf[ \cos\lf( \f{2\pi x}{L_\perp} + 1.4 \ri) + \cos\lf( \f{2\pi y}{L_\perp} + 0.5 \ri) \ri] \nonumber\\
  A_\para(x, y) &=& \f{\delta B_{\perp0}}{2}\lf( \f{L_\perp}{2\pi} \ri) \lf[ \f{1}{2}\cos\lf( \f{4\pi x}{L_\perp} + 2.3 \ri) + \cos\lf( \f{2\pi y}{L_\perp} + 4.1 \ri) \ri],
  \label{e:OT2 initial condition}
\end{eqnarray}
where $\delta u_0$ and $\delta B_{\perp0}$ represent the initial $E\times B$ drift speed and the initial amplitude of $\delta \bm{B}_\perp = \nbl_\perp A_\para\times\zhat$, respectively.
Here we have used unnormalized units.
In the following test, we choose the initial amplitude so that $\delta u_0 = \delta B_{\perp0}/\sqrt{4\pi m_\rmi n_\rmi}$.
Below we conduct the simulation in two characteristic spatial regions, the MHD inertial range and the ion kinetic range. 

\begin{figure}[htpb]
  \begin{center}
    \includegraphics*[width=1.0\textwidth]{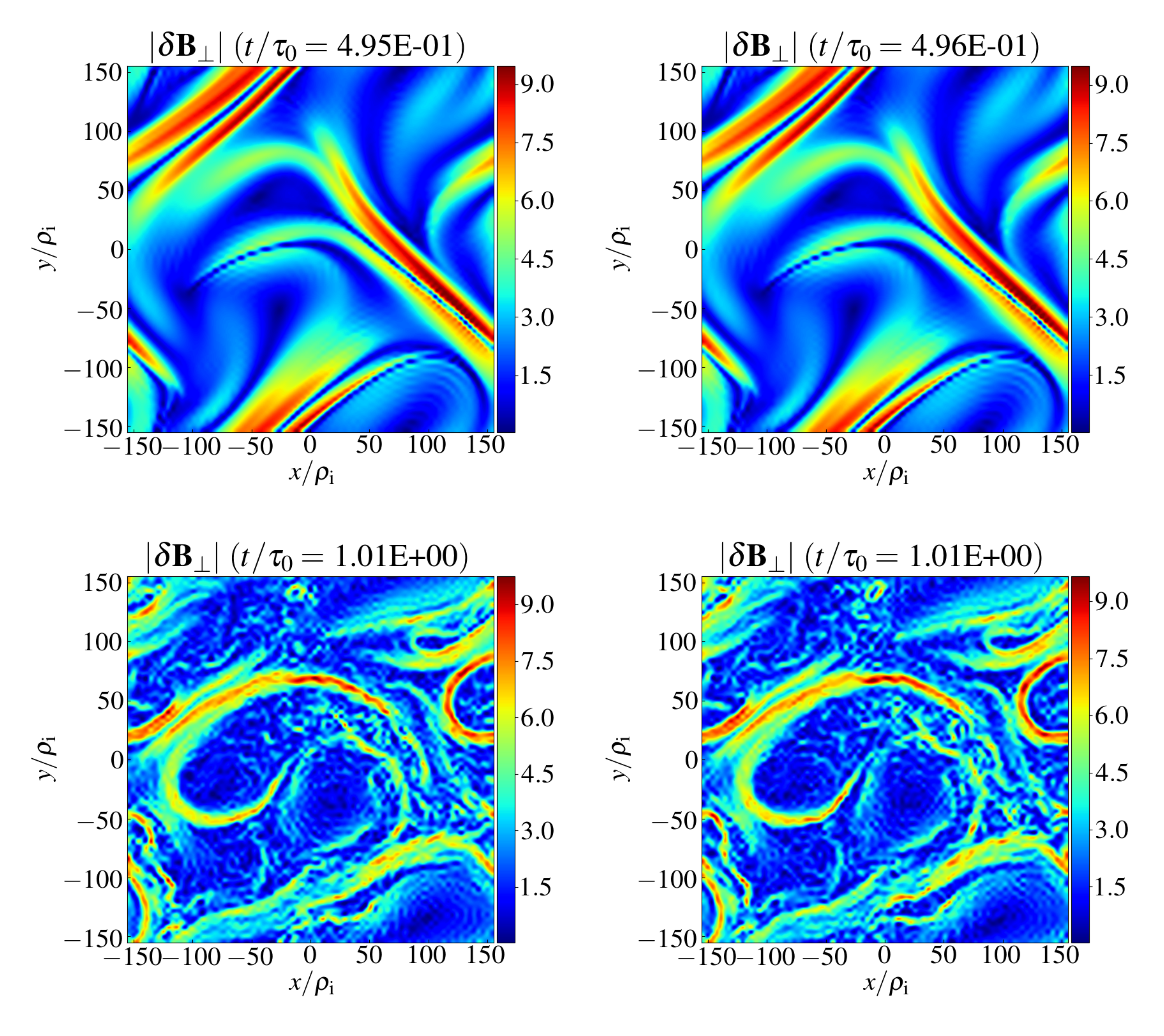}
  \end{center}
  \caption{Spatial profiles of $|\delta \bm{B}_\perp|$ for FGK (\textit{left}) and GKI/ITEF (\textit{right}) simulations at the current sheet formation phase (\textit{top}) and the turbulence phase (\textit{bottom}).}
  \label{f:OT2 field inertial}
\end{figure}

\begin{figure}[htpb]
  \begin{center}
    \includegraphics*[width=0.7\textwidth]{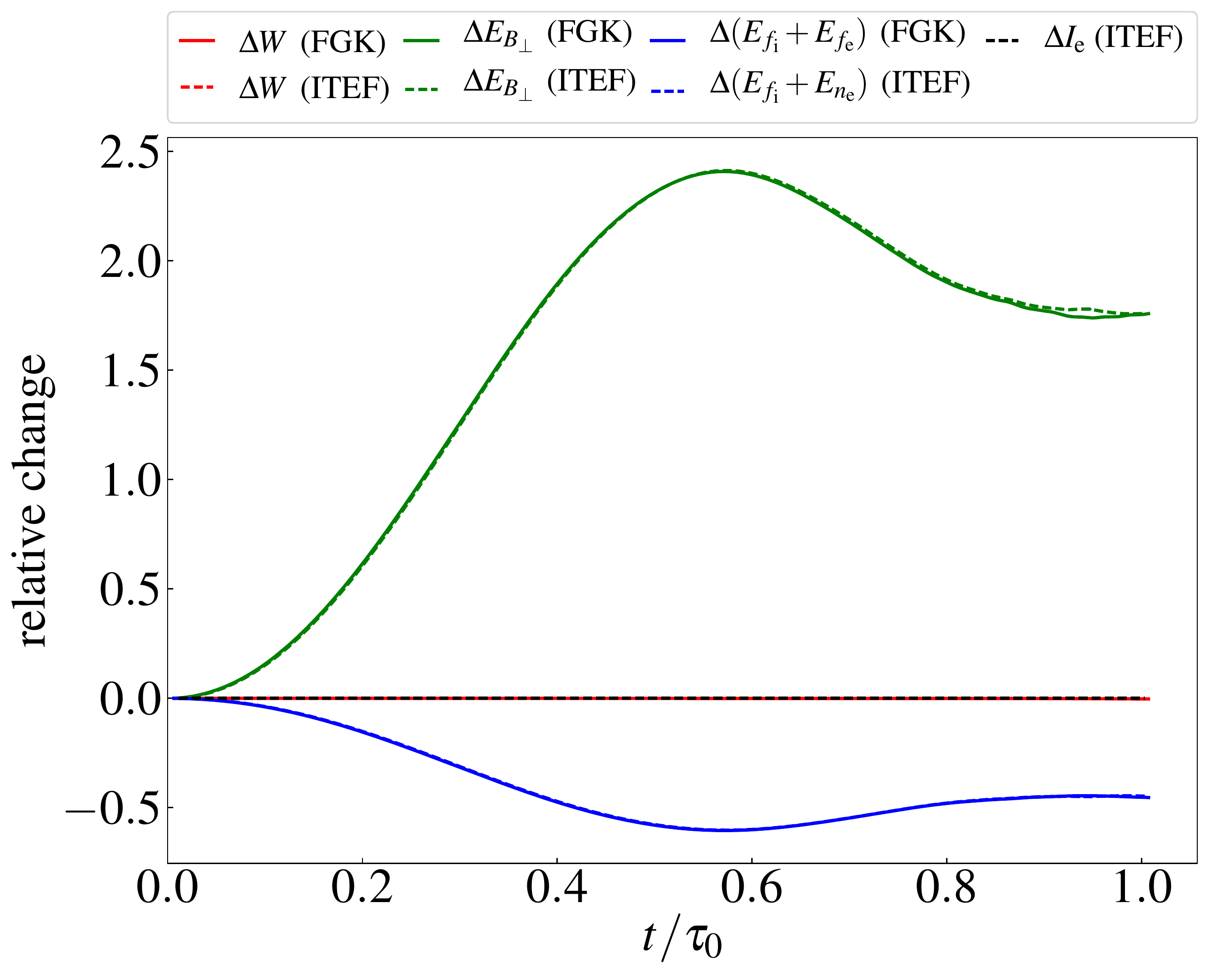}
  \end{center}
  \caption{Relative change of total energy and its components from the initial values for the FGK (\textit{solid lines}) and GKI/ITEF (\textit{broken lines}) simulations, and the two--dimensional invariant $I_\rme$ for GKI/ITEF (\textit{black broken line}).}
  \label{f:OT2 energy inertial}
\end{figure}

\begin{figure}[htpb]
  \begin{center}
    \includegraphics*[width=1.0\textwidth]{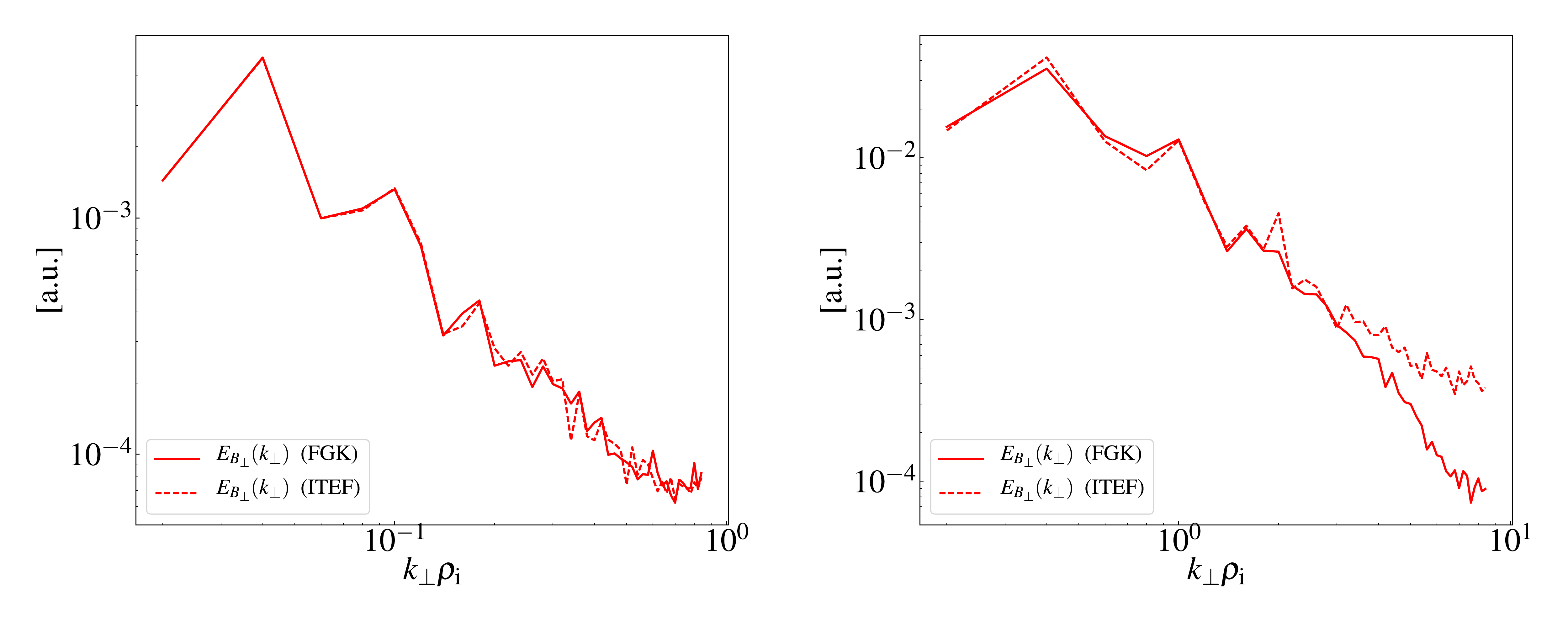}
  \end{center}
  \caption{Magnetic energy spectrum for FGK (\textit{solid lines}) and GKI/ITEF (\textit{broken lines}) in MHD inertial range (\textit{left}) and ion kinetic range (\textit{right}).}
  \label{f:OT2 spectra}
\end{figure}

For the MHD inertial range simulation, the domain is set to $L_\perp = 50\pi\rho_\rmi$ with the number of grid points $(n_x, n_y, n_\lambda, n_E) = (128, 128, 8, 16)$.
This gives a wave number range of $0.02 \le k_\perp\rho_\rmi \le 0.84$.
The plasma parameters are set to $\beta_\rmi = 1$, $\tau = 1$, and $Z = 1$.
A weak ion collisionality of $\nu_\rmi = 0.01\tau_0^{-1}$ with $\tau_0 = L_\perp/\delta u_0$ is imposed while the electrons are set to collisionless in the FGK simulation.
Figure~\ref{f:OT2 field inertial} shows the snapshots of $|\delta \bm{B}_\perp|$ at the current sheet formation phase and the turbulence phase.
At the current sheet formation phase, the profiles are almost identical between FGK and GK/ITEF. 
At the turbulent phase, while minor differences are seen in the small scale structures, the large scale structures such as the shape of the filaments are consistent.
Figure~\ref{f:OT2 energy inertial} shows the time evolution of the relative change in the total energy and its components.
There is good agreement between the GKI/ITEF and FGK simulations.
Since the collisionality employed here is tiny, the total energy $W$ and the 2D invariant $I_\rme$ are almost conserved.
We also conducted a simulation with zero collisionality, and found that the relative errors of $W$ and $I_\rme$ are within the order of $10^{-5}$ and $10^{-7}$, respectively, until the current sheet formation phase, $t/\tau_0 \simeq 0.5$ (after this time, collision is necessary to dissipate the small scale energy properly).
Figure~\ref{f:OT2 spectra} (\textit{left}) shows the spectrum of the magnetic energy at the turbulent phase.
Both spectra from the GKI/ITEF and FGK simulations agree very well over the entire wave number domain.

Next we show the result of the Orszag-Tang problem for the ion kinetic range.
We set the simulation domain to $L_\perp = 5\pi\rho_\rmi$ with the same number of grids used for the MHD inertial range simulation.
The corresponding wave number range is $0.2 \le k_\perp\rho_\rmi \le 8.4$ which spans the transition from the inertial to the kinetic range.
Figure~\ref{f:OT2 spectra} (\textit{right}) shows the magnetic energy spectrum at the turbulent phase.
The slope of the spectrum from the GKI/ITEF simulation is shallower than that from the FGK simulation.
This observation is consistent with the recent report on the comparison of FGK and a hybrid model composed of full kinetic ion and isothermal electron fluid model~\cite{Groselj2017}, although the plasma parameter setting is different.
The discrepancy may be attributed to the perpendicular electron damping, which was found to be effective in the ion kinetic region~\cite{Li2016}. 

\begin{figure}[htpb]
  \begin{center}
    \includegraphics*[width=1.0\textwidth]{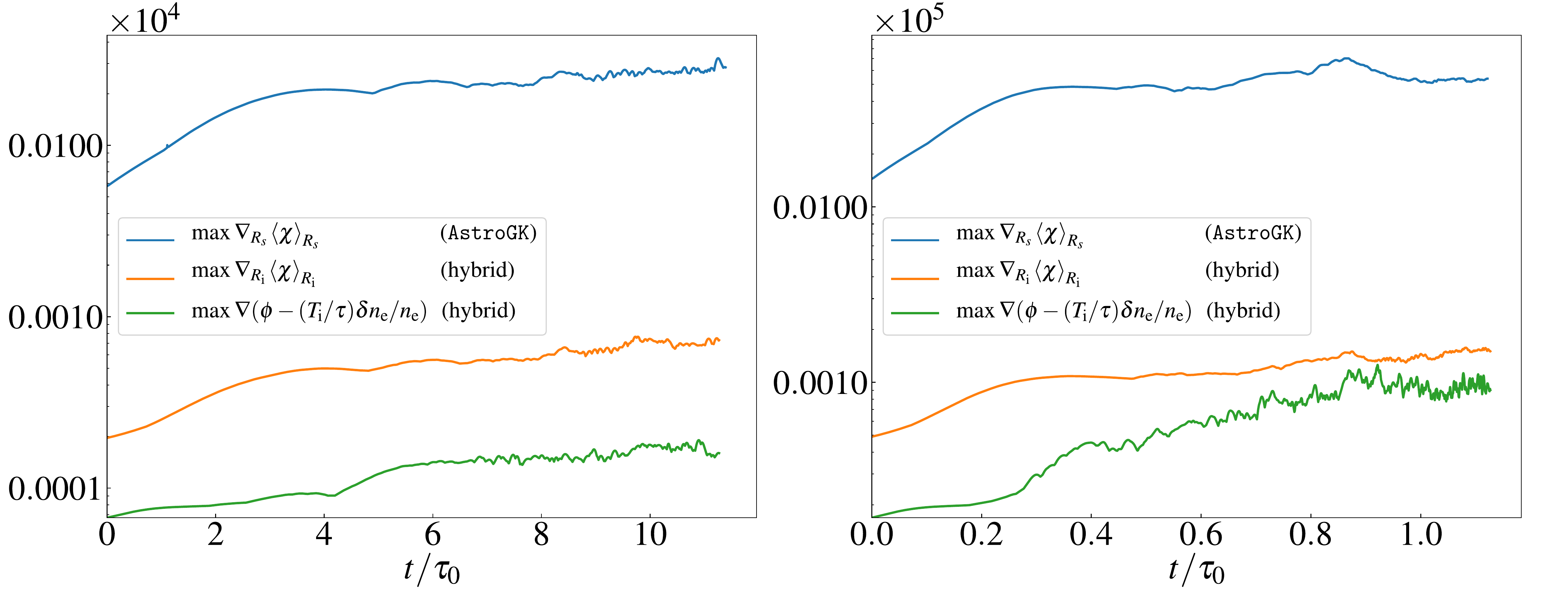}
  \end{center}
  \caption{Maximum advection speed for \agk\ and the hybrid code in MHD inertial range (\textit{left}) and ion kinetic range (\textit{right}) simulations.}
  \label{f:OT2 uperp}
\end{figure}

The improvement of the CFL condition estimated in Section~\ref{s:CFL improvement} is confirmed by comparison of \agk\ and the hybrid code.
Figure~\ref{f:OT2 uperp} shows the time evolution of the maximum advection speed for \agk\ and for the hybrid code in the inertial range and ion kinetic range simulations.
In both ranges, the maximum advection speed for GKI/ITEF is due to the ions, and the ITEF does not affect the CFL condition. 
On the other hand, the maximum advection speed for FGK (presumably dominated by the electron part) is approximately $\sqrt{m_\rmi/m_\rme}$ times greater than GKI/ITEF, which is consistent with the estimate in Section~\ref{s:CFL improvement}.
Using the series data of the maximum convection speed, we calculate the maximum timestep size as a function of time by interpolation.
Then, we estimate the minimum number of timestep necessary to reach $t/\tau_0 = 1$, which is 359681 for \agk\ and 8271 for the hybrid code.
On the other hand, the averaged computation time per a timestep is 0.756 CPU minutes for \agk\ and 0.430 CPU minutes for the hybrid code at ARCUS (Phase B) in the University of Oxford.
Multiplying the computational time per a timestep by the minimum step number, we estimate 4531.98 CPU hours for \agk\ and 59.23 CPU hours for the hybrid code. 
The improvement of CPU hours is 76.46, which is close to the ideal improvement $2\sqrt{m_\rmi/m_\rme} \sim 85.7$.
The slight fall-off from the ideal improvement is the parallel performance downtick due to the small grid number.
We confirmed that the improvement becomes exactly $2\sqrt{m_\rmi/m_\rme}$ when we increase the velocity grid number as $(n_\lambda, n_E) = (8, 16) \to (16, 32)$. 


\section{Conclusions}\label{s:conclusion}
A new hybrid simulation code for the GKI/ITEF model~\cite{Schekochihin2009ApJ} has been developed by extending \agk, an Eulerian $\delta f$ gyrokinetics code specialized to a slab geometry~\cite{Numata2010JCP}.
We have implemented an algorithm for implicitly solving the coupled system of ITEF and Maxwell's equations together with the ion gyrokinetic equation.
The linear terms are treated by the second-order compact finite difference method while the nonlinear terms are treated by the third-order Adams--Bashforth method.
Although the matrix to be inverted for solving the GKI/ITEF-Maxwell's system of equations is more complicated than that for \agk, the computational cost for the inversion is unchanged.
Therefore, the hybrid code retains the excellent parallel performance of \agk.
Since the fast electron timescale is eliminated in the hybrid model, the CFL condition for the explicitly handled nonlinear terms is dramatically less restrictive than for \agk.
We have estimated and confirmed that the hybrid code runs $2\sqrt{m_\rmi/m_\rme} \sim 100$ times faster than \agk\ with a single ion species and gyrokinetic electrons. 

We also have presented linear and nonlinear tests for code verification.
The linear test reproduces the theoretical predictions showing that the implicit algorithm for ITEF is correctly implemented.
For the nonlinear test, we conducted the 2D Orszag--Tang vortex problem for spatial scales both larger and smaller than the ion Larmor radius. 
In the former range, the hybrid code gives a result identical to \agk\ (and also reduced MHD) as theoretically predicted.
On the other hand, the power spectrum of GKI/ITEF is shallower than that of FGK at scales smaller than the ion Larmor radius.
This is consistent with a recent numerical study comparing FGK and a full kinetic ion and isothermal electron fluid hybrid model~\cite{Groselj2017}.

One possible application of the hybrid code is a study on the parametric dependence of ion and electron turbulent heating,
which is one of the most important problems in both inner and extra solar systems.
At the ion Larmor scale, the energy cascaded from the large scales splits into ion entropy fluctuations and KAWs. 
It is theoretically predicted that ion entropy fluctuations and KAWs are independently cascaded to smaller scale~\cite{Schekochihin2009ApJ}. 
The former leads to ion heating while the latter leads to the electron heating.
Therefore, the partition of heating by the dissipation of Alfv\'{e}nic turbulence is determined at the ion Larmor scale, and we do not need to resolve the electron kinetic scale.
Since electron kinetics effects are eliminated from the model, we measure only the ion heating occurring around the ion Larmor scale; electron heating may be estimated by assuming that all of the injected energy not dissipated at these scales will ultimately end up heating the electrons.
Since the investigation of heating requires high resolution in velocity space, it is computationally cumbersome to scan plasma parameters, e.g., $\beta_\rmi$ and $\tau$.
However, with the improved computational efficiency of the hybrid code, such a parameter scan should be feasible. 

Another interesting application of the hybrid code is to observe the phase space cascade in \emph{3D electromagnetic} turbulence~\cite{Schekochihin2008,Schekochihin2009ApJ, Plunk2010}. 
Whereas the phase space cascade was observed for electrostatic turbulence by \agk\ in a restricted 4D phase space with fine space and velocity grids~\cite{Tatsuno2009,Tatsuno2010,Plunk2011,Tatsuno2012}, e.g., $(n_x, n_y, n_\lambda, n_E) = (256, 256, 192, 96)$~\cite{Tatsuno2012}, it is unrealistic to conduct similar high resolution simulations for a 5D electromagnetic case with a FGK code.
Again, we expect that such simulations are feasible with the present hybrid code at reasonable computational cost. 

\section*{Acknowledgements}
This work was supported by STFC grant ST/N000919/1.  
We thank A. A. Schekochihin and D. Gro{\v{s}}elj for useful discussions.
The authors also acknowledge the use of ARCHER through the Plasma HEC Consortium EPSRC grant number EP/L000237/1 under the projects e281-gs2 and the use of the University of Oxford Advanced Research Computing (ARC) facility.


\bibliographystyle{elsarticle-num} 
\bibliography{references}





\end{document}